\documentclass[reprint,aps]{revtex4-2}
\usepackage{bm}
\usepackage{amsmath}
\usepackage{amssymb}
\usepackage{graphicx}
\usepackage[colorlinks=true,linkcolor=blue,citecolor=blue,urlcolor=blue]{hyperref}
\usepackage{siunitx}
\usepackage{color}
\usepackage{braket}

\begin{document}

\title{Layer Pseudospin Magnetism in Transition-Metal-Dichalcogenide Double-Moir\'es}

\author{Yongxin Zeng}
\thanks{These authors contributed equally to this work.}
\author{Nemin Wei}
\thanks{These authors contributed equally to this work.}
\author{Allan~H.~MacDonald}
\affiliation{Department of Physics, University of Texas at Austin, Austin TX 78712}

\date{\today} 

\begin{abstract}
Spontaneous order of layer pseudospins in two-dimensional bilayers is common in quantum Hall systems,
where it is responsible for hysteretic responses to gate fields in states with Ising order,
and giant drag voltages in states with XY (spontaneous inter-layer phase coherence) order. 
In this article we predict that layer pseudospin order will also occur in double-moir\'e 
strongly correlated two-dimensional electron systems.  
We comment on similarities and differences in the competition between the two types of order
in quantum Hall and double-moir\'e systems, and relate our findings to previous work on 
Falicov-Kimball models of electronic ferroelectrics.
\end{abstract}

\maketitle

\section{Introduction}
Bilayer two-dimensional electron systems possess a {\em which layer} degree
of freedom that is conveniently regarded as an artificial pseudospin.  When the two layers are electrically 
isolated, conservation of their electron number difference is manifested by  
invariance under global rotations about the $\hat{z}$-direction in layer pseudospin 
space, like the spins of a lattice XXZ model which in two-dimensions can have 
Ising or XY Kosterlitz-Thouless order depending on model parameters.
In the case of bilayers in the quantum Hall regime \cite{eisenstein2004bose,eisenstein2014exciton}, it has been established that at some Landau level filling factors $\nu$,the layer pseudospins have XY order.
Layer pseudospin order is especially robust near $\nu=1$, where the ordered state can be viewed as an 
exciton-condensate of electrons in the lowest Landau level of one layer and the holes in the lowest 
Landau level of the other layer, and is responsible for fantastic electrical anomalies including 
large transport drag signals and dissipationless counterflow transport.  

\begin{figure}
    \centering
    \includegraphics[width=\linewidth]{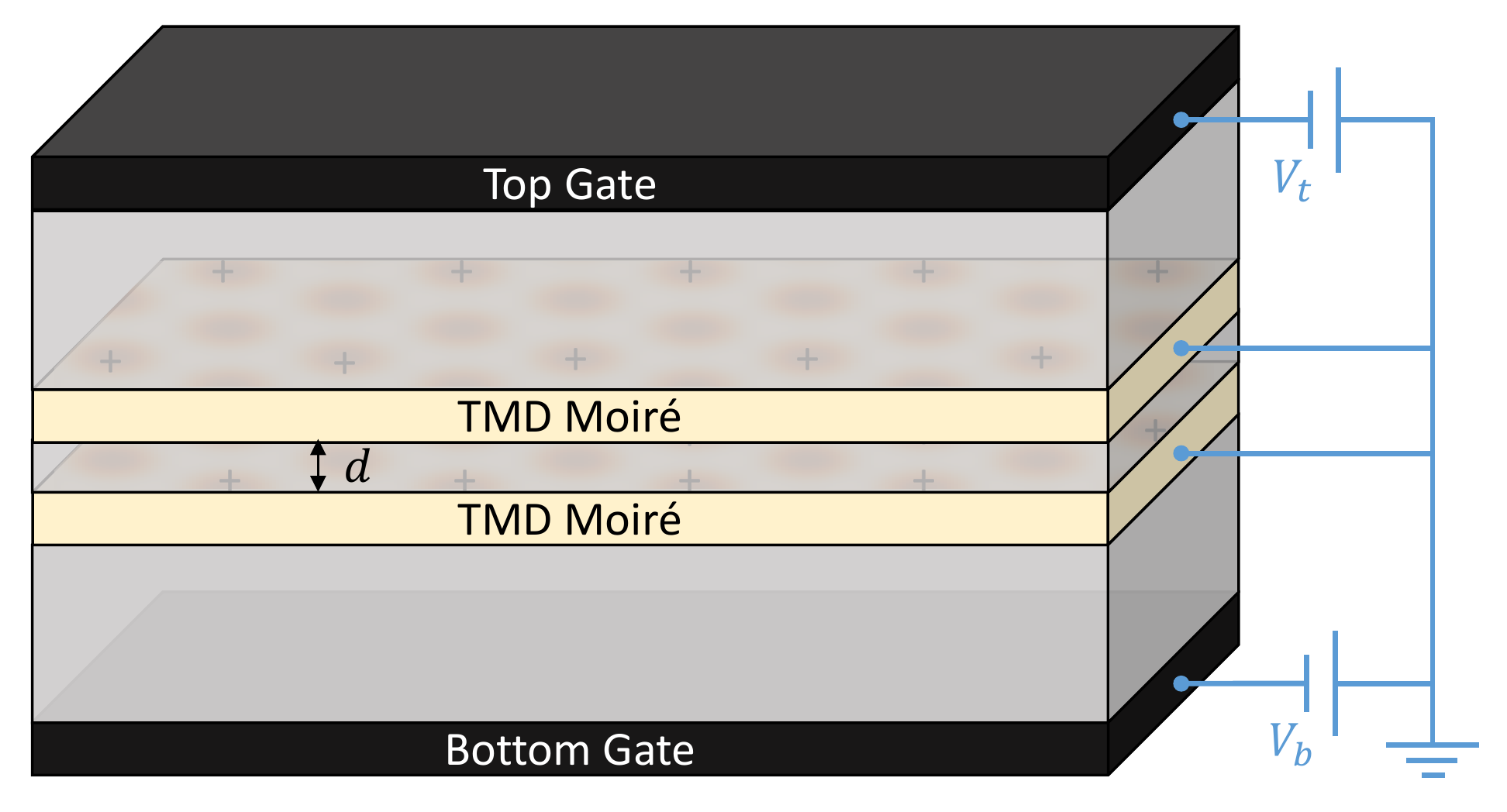}
    \caption{A dual-gated transition metal dichalcogenide (TMD)
    double-moir\'e system.  Each moir\'e layer has a periodic lateral modulation.
    The two moir\'e layers are assumed to be identical and 
    perfectly aligned horizontally. Negative voltages are applied at the top and bottom gates to induce positive charge carriers (holes) in the moir\'e layers, both of which are grounded.  A difference between the top and bottom gate voltages produces a vertical electric field that creates an electric potential difference between the two moir\'e layers that can be varied at fixed carrier density.
    Dielectric tunnel barriers (gray regions) are inserted between the layers to suppress interlayer tunneling.}
    \label{fig:device}
\end{figure}

In recent years, experimenters have developed moir\'e superlattices \cite{cao2018correlated,cao2018unconventional,regan2020mott,tang2020simulation,xu2020correlated}, two-dimensional 
semiconductors or semimetal bilayers in which a moir\'e pattern has formed, as an attractive
platform for studies of highly tunable strong correlation physics.
In this article we propose that the bilayer counterflow superfluid states discovered first 
in quantum Hall systems \cite{spielman2000resonantly,kellogg2004vanishing,tutuc2004counterflow} also occur at zero magnetic 
field in double moir\'es -- systems with two moir\'e superlattices separated by an insulating layer
as illustrated schematically in Fig.~\ref{fig:device}.
As in the quantum Hall case, counter-flow superfluids are most stable near $\nu=1$, 
where $\nu$ in this case is the number of carriers per moir\'e lattice site.
In the double-moir\'e case, the spontaneous coherence states 
compete with a series of broken translational symmetry exciton crystal
states that reduce to lattice gas states in 
the long-moir\'e period limit and are responsible for hysteretic response of the layer polarization to 
externally applied displacement fields.  We find that the crystalline states prevail at small fields
and at small twist angles, whereas the layer coherent states are more common at larger displacement fields before the system become fully layer polarized. For the honeycomb (bipartite) lattice case, 
this spontaneous coherence state can also be viewed as spin-flop states of the layer pseudospin. For the triangluar (non-bipartite) lattice, layer-magnetic frustration forces 
coherent states to break the translational symmetry and become supersolid phases of excitons. 

\section{TMD double-moir\'e systems} \label{sec:model}
The low-energy electronic physics of long-period moir\'e superlattices is accurately described by continuum models \cite{bistritzer2011moire,wu2018hubbard,angeli2021gamma}. In the case of p-type group-VI transition metal dichalcogenide semiconductor (TMD) heterobilayers \cite{wu2018hubbard}, there is only one low-energy orbital 
state per spin, and the moir\'e pattern acts like a periodic modulation potential.
The single-particle Hamiltonian of valence band holes is therefore 
\begin{equation} \label{eq:H0}
H_0 = -\frac{\hbar^2 \bm k^2}{2m} + \Delta(\bm r),
\end{equation}
where $m$ is the effective mass and $\Delta(\bm r)$ is the moir\'e modulation potential. 
The moir\'e potential extrema form a triangular lattice with period $a_M$, the moir\'e lattice constant.

The situation is different for AA stacked 
TMD homobilayers \cite{xian2021realization,angeli2021gamma} because  strong interlayer hybridization 
leads to an emergent $C_2$ symmetry in the moir\'e potential. 
The valence band maximum for most TMD homobilayers lies at the $\Gamma$ point and is spin-degenerate. The low-energy physics of valence band holes is described by the same Hamiltonian as in the 
heterobilayer case \eqref{eq:H0}, except that the moir\'e potential has higher symmetry and the potential maxima form a honeycomb lattice.

In this work we consider two moir\'e TMD bilayers separated by a few-layer hexagonal boron nitride (hBN) tunnel barrier, as shown in Fig.~\ref{fig:device}. We assume the two moir\'es are composed of 
the same materials, that they have the same moir\'e periods, and that 
they are perfectly aligned both rotationally and translationally. Later we will discuss the experimental 
relevance of these assumptions, and the robustness of our results in the presence of a small lateral displacement between the two moir\'es. The double-moir\'e system is described by the Hamiltonian
\begin{equation}
H = H_0 \otimes \tau_0 - \frac{V_z}{2} \tau_z + H_1,
\end{equation}
where $\tau_0$ and $\tau_z$ are the identity matrix and Pauli-$z$ matrix in layer pseudospin space, $V_z$ is the electric potential difference between the two layers produced by a perpendicular electric field, and $H_1$ is the Coulomb interaction between electrons:
\begin{equation}\label{eq:coulomb}
H_1 = \frac{1}{2A} \sum_{ll'} \sum_{\bm k \bm k' \bm q} V_{ll'}(q) a_{\bm k + \bm q, l}^{\dagger} a_{\bm k' - \bm q, l'}^{\dagger} a_{\bm k', l'} a_{\bm k, l}.
\end{equation}
Here $A$ is the area of the two-dimensional system and $l,l'$ are the layer indices. To emphasize the pseudospin analogy we label the two layers as $\uparrow, \downarrow$. The intralayer and interlayer Coulomb interactions are
\begin{equation} \label{eq:V_q}
V_{ll'}(q) = 
\begin{cases}
2\pi e^2/\epsilon q, & l=l', \\
(2\pi e^2/\epsilon q) e^{-qd}, & l\ne l',
\end{cases}
\end{equation}
where $\epsilon$ is the effective dielectric constant and $d$ is the effective layer separation between the two moir\'es. In this article we focus on layer pseudospin magnetism and neglect electron spin (valley) degree of freedom by assuming that the energy scale associated with spin ordering is much lower than that associated with layer pseudospin ordering. We will later discuss the validity of this assumption and briefly explain how spin ordering may affect our results.

\section{Effective lattice models}
The continuum model we study reduces in certain limits
to lattice models that have been extensively studied in the literature and have properties that 
are well understood.  Below we will refer to these lattice models to provide intuition on the 
physics behind our continuum model.  Lattice models are 
most relevant in the limit of long moir\'e periods in which 
holes are strongly localized at the moir\'e potential maxima, which can form triangular or honeycomb lattices as we have explained.

The single-particle physics in the lattice limit is accurately described by a tight-binding model \cite{wu2018hubbard}
\begin{equation} \label{H0_lattice}
\mathcal{H}_0 = -\sum_{ij,l} t_{ij,l} c_{il}^{\dagger} c_{jl} - \frac{V_z}{2} \sum_i (n_{i\uparrow} - n_{i\downarrow}),
\end{equation}
where $c_{il}^{\dagger}$ is the creation operator of a hole in the localized Wannier orbital at site $i$ in layer $l$, $n_{il}$ is the number operator, $\bm R_i$ is the position of site $i$, and $t_{ij,l}$ is a hopping parameter in layer $l$ that decreases rapidly with the distance between sites $|\bm R_i - \bm R_j|$. In the strong moir\'e modulation limit, the interaction Hamiltonian takes the generalized Hubbard form
\begin{equation} \label{H1_lattice}
\mathcal{H}_1 = \frac 12 \sum_{ij,ll'} U_{ij,ll'} c_{il}^{\dagger} c_{jl'}^{\dagger} c_{jl'} c_{il},
\end{equation}
where $U_{ij,ll'} = U_{ll'}(|\bm R_i - \bm R_j|)>0$ is the interaction energy between site $i$ in layer $l$ and site $j$ in layer $l'$. The value of $U$ can in principle be calculated \cite{wu2018hubbard} by projecting Coulomb repulsion onto the localized Wannier orbitals, but at large distance it takes a simple Coulomb form
\begin{equation}
\label{eq:Coulombtail}
U_{ll'}(r) \approx
\begin{cases}
e^2/\epsilon r, & l=l',\\
e^2/\epsilon \sqrt{r^2+d^2}, & l\ne l'.
\end{cases}
\end{equation}
Eq.~\ref{eq:Coulombtail} applies when the
spatial extent of the Wannier orbitals is negligible compared to the inter-site distance. We notice that the lattice Hamiltonian \eqref{H0_lattice}-\eqref{H1_lattice} has the same form as the extended Falicov-Kimball model, which has been extensively studied in the literature  \cite{kunevs2015excitonic,batista2002electronic,portengen1996theory,batista2004intermediate,farkasovsky2008hartree,kaneko2013exact}, except that the interactions and hopping terms are generalized beyond the on-site and nearest-neighbor contributions that are normally retained.

The pseudospin analogy is most transparent in the strong interaction limit when we assume that 
the filling factor is such that on average one hole is present at each site. In the language of band filling, the filling factor $\nu=1$ in the triangular lattice case and $\nu=2$ in the honeycomb lattice case. Since the on-site repulsion $U_{\uparrow\downarrow}(0)$ is much stronger than the repulsion between different sites when the two moir\'es are horizontally aligned, the low-energy subspace of the system consists
in thew weak inter-site hopping limit of states with only one hole at each site in one of the two layers. The {\em which layer} degree of freedom then acts as a localized layer pseudospin at each lattice site that interacts with neighboring pseudospins.  In the rest of the paper we study the 
consequences of these interactions for pseudospin magnetic order.

If we start from the lattice Hamiltonian \eqref{H0_lattice} and \eqref{H1_lattice}, 
treat the hopping terms as perturbations, and expand the Hamiltonian in the low-energy subspace where each site $i$ is occupied once, we obtain (up to order $O(t^2/U)$) the XXZ spin model:
\begin{equation} \label{eq:H_XXZ}
\mathcal{H}_{\rm XXZ} = \sum_{i<j} \left[ J_{ij}^z \tau_i^z \tau_j^z + J_{ij}^{\perp} (\tau_i^x \tau_j^x + \tau_i^y \tau_j^y) \right] - \frac{V_z}{2} \sum_i \tau_i^z,
\end{equation}
with the coupling parameters
\begin{equation} \label{eq:J_ij}
J_{ij}^z = \frac{U_{ij,\uparrow\uparrow} - U_{ij,\uparrow\downarrow}}{2} + \frac{t_{ij,\uparrow}^2 + t_{ij,\downarrow}^2}{2U_{\uparrow\downarrow}(0)},\quad
J_{ij}^{\perp} = \frac{t_{ij,\uparrow}t_{ij,\downarrow}}{U_{\uparrow\downarrow}(0)}.
\end{equation}
Eq.~\eqref{eq:J_ij} shows that the pseudospin couplings are easy-axis antiferromagnetic: $J_{ij}^z > J_{ij}^{\perp} > 0$. Besides the usual $t^2/U$ terms, the difference between intralayer and interlayer Coulomb repulsions gives rise to an extra term in $J_{ij}^z$. Since the hopping parameter $t$ decays exponentially with distance \cite{wu2018hubbard}, at large distance the Coulomb term dominates and the pseudospin coupling is of dipolar form
\begin{equation}
J_{ij}^z \approx \frac{e^2 d^2}{4\epsilon |\bm R_i-\bm R_j|^3}.
\end{equation}
As we will see later, the long-range nature of pseudospin couplings plays an important role in the rich phases the system displays.

\section{Mean-field phase diagrams}
\begin{figure*}
   \includegraphics[width=2\columnwidth]{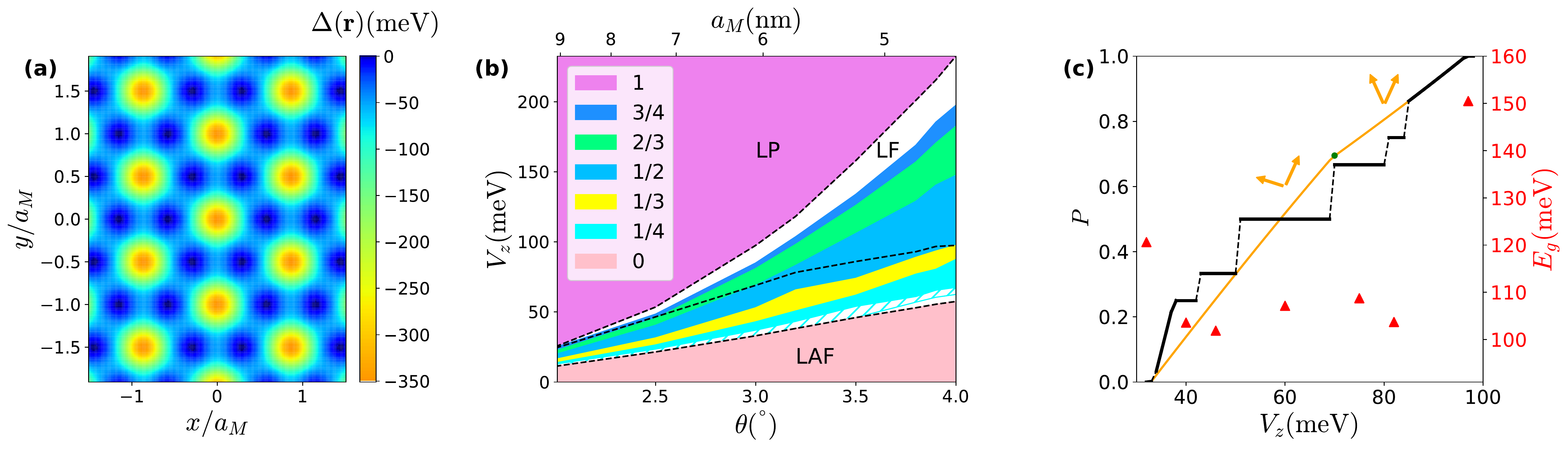}
    \caption{ (a) Model modulation potential for a moir\'e TMD homobilayer.  Valence band holes are localized at the potential maxima (shifted to zero energy here), which in this case form a honeycomb lattice. 
    This modulation potential describes\cite{angeli2021gamma} holes in WS$_2$ homobilayers. 
    (b) Phase diagram of double WS$_2$ homobilayer moir\'es at hole-filling factor $\nu=2$ (one hole per site) in the plane of displacement field $V_z$ and twist angle $\theta$.  When the possibility of broken translational symmetry is 
    discarded we find four distinct phases separated by continuous phases marked by dashed lines; a layer antiferromagnet (LAF) at weak $V_z$, and layer-polarized (LP) state at large $V_z$, and two distinct layer-coherent states
    at intermediate values of $V_z$. The layer-coherent state at larger $V_z$ is analogous to the spin-flop state in magnetic systems, and we therefore call it the layer-flop (LF) state. The colored stripes at intermediate $V_z$ represent dipole crystal states with various different layer polarizations ($P=(n_{\uparrow}-n_{\downarrow})/(n_{\uparrow}+n_{\downarrow})$), that appear only 
    when translational symmetry breaking is permitted. The hatched region is a supersolid state with both interlayer phase coherence and broken translational symmetry. and the remaining white regions are layer-coherent states that preserve translational symmetry. (c) The layer polarization $P$ as a function of displacement field $V_z$ at fixed twist angle $\theta=\SI{3}{\degree}$. The orange curve shows the result obtained for layer-coherent states when translational symmetry breaking is not allowed. The transition point between the two distinct layer-coherent states is marked by the green dot, and the difference between their layer pseudospin configurations is schematically indicated by arrows. 
    The black curve illustrates the layer-polarization pleateaus obtained when translational 
    symmetry breaking is allowed. 
    The red triangles show the intra-layer charge excitation gaps in the majority layer of the layer-incoherent states,
    which we expect to be relevant to transport properties. 
}
    \label{fig:gamma}
\end{figure*}

We study the pseudospin order of double-moir\'e systems by projecting the 
Coulomb interaction \eqref{eq:coulomb} onto the highest moir\'e bands that are relevant -- one band for triangular lattice systems and two bands for honeycomb lattice systems -- and then approximating interaction effects using
Hartree-Fock mean-field theory. We construct the phase diagrams by identifying changes in the symmetries 
of the lowest-energy self-consistent solutions
as the twist angle and displacement field tuning parameters are varied. 
The numerical calculations are performed for two different moir\'e modulation
potentials which illustrate the honeycomb-lattice (homobilayer) and triangular-lattice (heterobilayer) cases. 
The moir\'e potentials for the two systems are plotted in Figs.~\ref{fig:gamma}(a) and \ref{fig:K}(a) respectively.
The derivation of the projected Hartree-Fock equations is detailed in Appendix~\ref{app:hf}, and the continuum model parameters for the two systems are specified in Appendix~\ref{app:parameters}.
Other system parameters include the dielectric constant $\epsilon=6$ and interlayer distance $d=\SI{2}{nm}$.

We start with the honeycomb lattice systems as they are expected to be simpler due to the lack of geometric frustration of near-neighbor antiferromagnetic couplings. Fig.~\ref{fig:gamma}(b) shows the mean-field phase diagram for WS$_2$ homobilayers at filling factor $\nu=2$ (two holes per unit cell, one hole per site) {\it vs.} twist angle $\theta$ and displacement field $V_z$.  For one-hole per site, the low energy states of insulators may be mapped to those of honeycomb lattices 
with a layer pseudospin degree-of-freedom.  If we limit our search for the ground state to states in which translational symmetry is not spontaneously broken, the phase diagram (see the dashed lines in Fig.~\ref{fig:gamma}(b)) is similar to that of the square-lattice XXZ model \cite{batista2002electronic}. When the displacement field is absent,
the two moir\'e layers have identical potentials and the system forms a layer-antiferromagnet (LAF) in which holes at neighboring sites are localized in different layers, breaking sublattice symmetry. At large field $V_z$ the system is fully layer-polarized (LP) and all holes move to the moir\'e layer with lower electric potential. 
At intermediate $V_z$ the system forms a layer-coherent state. In the language of layer pseudospins localized at moir\'e lattice sites, the pseudospins develop an in-plane component at intermediate $V_z$. The in-plane components of the pseudospins on the two sublattices point in opposite directions due to the antiferromagnetic coupling. 

In our mean-field results we find two different layer-coherent states. At large $V_z$ near the LP state, the pseudospins on both sublattices have the same $z$-component and equal but opposite in-plane components. This is analogous to the spin-flop phase of canted antiferromagnets in a magnetic field, and we call it the layer-flop (LF) state. In the other layer-coherent state at smaller displacement field, the $z$-components of neighboring pseudospins are different \cite{holtschneider2007biconical,yamashita1972field,matsuda1970off,liu1973quantum,bruce1975coupled}. 
The small $V_z$ state is stabilized by the long-range interaction $J_{ij}^z$ \cite{holtschneider2007biconical}. We find that the layer-coherent state regions in the phase diagram get dramatically wider as the twist angle increases. This is because the in-plane pseudospin coupling strength $J^{\perp}$ is proportional to the square of the hopping parameter $t$ between different sites (Eq.~\eqref{eq:J_ij}), which increases as the neighboring sites get closer at larger twist angles. Our mean-field results show that all four phases are connected by continuous phase transitions. The orange line in Fig.~\ref{fig:gamma}(c) shows the layer polarization (defined as $P=(n_{\uparrow}-n_{\downarrow})/(n_{\uparrow}+n_{\downarrow})$ where $n_{l}=\sum_{i}n_{il}$) as a function of $V_z$ at fixed $\theta=\SI{3}{\degree}$, together with the schematic illustration of the pseudospin orientations on neighboring sites that distinguish the two layer-coherent states.

The phase diagram becomes much more complex when we allow translational symmetry breaking in our calculations. 
The colored stripes in Fig.~\ref{fig:gamma}(b) show the low-energy states that emerge when we perform the calculations in $\sqrt{3}\times\sqrt{3}$ and $2\times 2$ supercells, with different colors representing states with different values of layer polarization. We see that inside the previously identified layer-coherent regions a series of lower-energy states
appear that break translational symmetry. 
Most of these states (solid filled regions) are dipole crystals without interlayer coherence (see Appendix~\ref{app:spin_plot} for details on the spatial distribution of layer pseudospins). Each of these dipole crystal states is stable over a 
finite range of displacement field $V_z$, and as shown in Fig.~\ref{fig:gamma}(c) the layer polarization curve (black) has a series of plateaus and discontinuous jumps. We also find supersolid states that break both layer-$U(1)$ and translational symmetry inside the hatched region in Fig.~\ref{fig:gamma}(b). The remaining white regions, mostly near the LP state region, are the layer-coherent states that do not break translational symmetry. Inside these layer-coherent state regions, the layer polarization varies continuously with the displacement field $V_z$. 

The phase diagram in Fig.~\ref{fig:gamma}(b) is not complete since more crystal states are expected to appear as we increase the maximum size of our supercells. These crystal states are stabilized by long-range dipolar interactions between different sites. The energy competition between these states is in general very complicated and sensitive to parameter choices. Nevertheless, the phase diagram contains two generic features. First, as the twist angle $\theta$ is reduced, the layer-coherent state regions rapidly narrow and nearly disappear at very small $\theta$.  This behavior is expected given that 
the in-plane pseudospin coupling $J^{\perp}$ decreases rapidly with $\theta$.  At 
small $\theta$ the system is well approximated by a lattice-gas model, in which classical charges are 
localized at lattice sites, and the polarization {\it vs.} $V_z$ curve approaches 
a {\em devil's staircase} structure \cite{yamamoto2012quantum,capogrosso2010optical,hubbard1978wigner,fisher1980infinitely,bak1982ising} with many
small closely spaced polarization jumps.  The ground state polarization $P$ is rational at all values of $V_z$.  Although 
large portions of the stability regions of the layer-coherent states are replaced by dipole crystal states that 
don't have interlayer coherence, the layer flop state remains the ground state in a region near the LP state and the width of this region increases with the twist angle. We can understand this behavior if we recognize that 
when the carriers are nearly polarized to one of the layers, the low-energy degrees of freedom are electron-hole excitations that move carriers from the majority layer to the minority layer, and the system is equivalent to a dilute gas of excitons. At low temperatures the excitons condense, establishing interlayer coherence. 
As we will see in the next section, this argument allows us to generalize some of our results beyond 
the perfectly-aligned limit of double-moir\'e systems on which we focus. 

In Fig.~\ref{fig:gamma}(c) we also plot the transport gap $E_g$ 
of the layer incoherent phases. 
In these states strong suppression of interlayer tunneling by the dielectric barriers 
implies that quasiparticles are localized in definite layers, which conduct independently.
The transport gap is defined as the charge gap between occupied and empty states localized in
the same layer since this is the quantity that controls thermally activated transport;
typically the values are similar in the majority and minority layers.
The charge gaps are therefore independent of $V_{z}$ within a given polarization plateau. 
The gap is maximized at large $V_z$ when the system is fully layer 
polarized and is significantly reduced at small $V_z$ where each layer is partially polarized.


\begin{figure}
   \includegraphics[width=1\columnwidth]{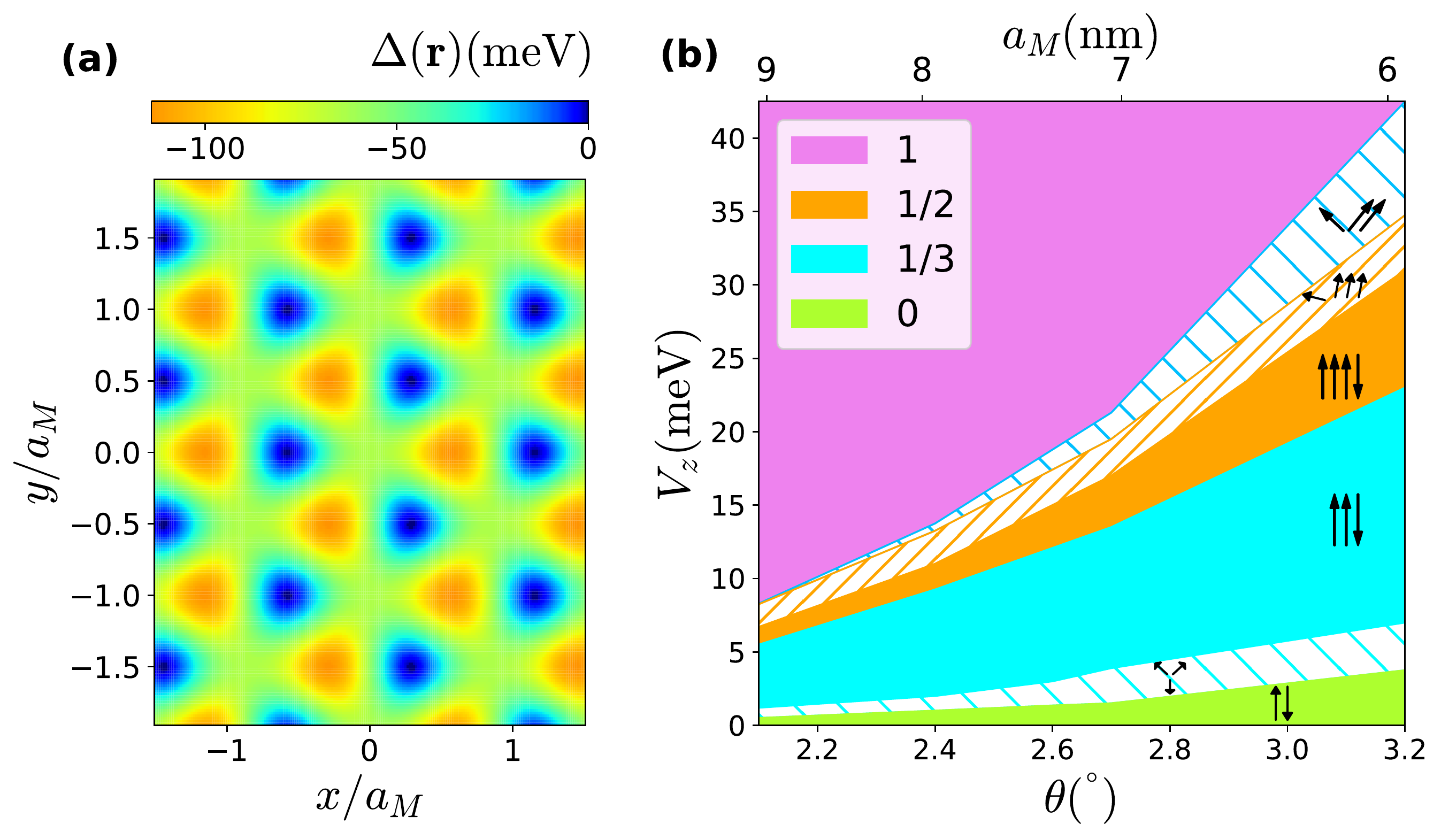}
    \caption{(a) The moir\'e potential for a TMD heterobilayer in which the potential maxima form a triangular lattice. The system parameters are adopted from Ref.~\cite{morales2021non}. 
    (b) The phase diagram of triangular-lattice double-moir\'{e}s at filling factor $\nu=1$ in the plane of displacement field $V_z$ and twist angle $\theta$. The solid filled regions represent states with no interlayer coherence and different colors represent different layer polarizations $P=(n_{\uparrow}-n_{\downarrow})/(n_{\uparrow}+n_{\downarrow})$. The hatched regions represent supersolid states that break both layer-$U(1)$ and translational symmetries. The arrows schematically show the pseudospin configurations of different states. 
}
    \label{fig:K}
\end{figure}

Next we turn to triangular lattice systems. Fig.~\ref{fig:K} shows the moir\'e potential and the phase diagram (at filling factor $\nu=1$) for a twisted TMD heterobilayer, with the same parameter choices as in Ref.~\cite{morales2021non}. 
We focus on the case of one-electron per site for both lattice types.  
Due to the geometric frustration of antiferromagnetic pseudospin couplings on triangular lattices, 
in this case all states in the phase diagram break translational symmetry except for the fully layer 
polarized states at large displacement field. 
For clarity we show in the phase diagram only states with $\sqrt{3}\times\sqrt{3}$ and $2\times 2$ supercells, although states with larger supercells are expected to be energetically preferred in part of the diagram due to the long-range dipolar interactions discussed previously. At zero displacement field the ground state is a stripe state in which all holes in a single stripe occupy one of the layers, while the holes in nearby stripes are localized in the other layer. 
At intermediate $V_z$ we find that dipole crystal states appear for both 
$\sqrt{3}\times\sqrt{3}$ and $2\times 2$ supercells.
In the phase diagram, we label these two states as $\uparrow\uparrow\downarrow$ and $\uparrow\uparrow\uparrow\downarrow$ respectively. In the remaining part of the phase diagram we find layer-coherent states that also break translational symmetry, {\it i.e.}, supersolid states (hatched regions). At small $V_z$ (near the stripe state)
we find supersolid states in which the three pseudospins within a supercell are oriented at approximately \SI{120}{\degree} with respect to each other.  As $V_z$ increases this state gradually deforms into the $\uparrow\uparrow\downarrow$ dipole crystal state. The other two supersolid states are located close to the LP state and 
have nonzero in-plane components and positive $z$ components for all pseudospins within a supercell. The pseudospin arrangements of all states (except the LP state) are schematically shown in Fig.~\ref{fig:K}. It is interesting to compare our phase diagram with those obtained in previous work on the triangular-lattice XXZ model \cite{yamamoto2014quantum,sellmann2015phase}. We see that although our results are in good agreement with those studies when restricted to $\sqrt{3}\times\sqrt{3}$ supercells, the long-range dipolar interactions in our case lead to states with larger supercell sizes in parts of the phase diagram.

\section{Discussion}
Insulating states are common at fractional layer polarizations of double-moir\'e systems. 
From our mean-field results we see that the physics of double-moir\'e Mott insulators
is in many ways similar to that of near-neighbor XXZ spin models under the influence of a 
magnetic field in the $\hat{z}$ direction, which has been studied in previous work ~\cite{batista2002electronic,yamamoto2014quantum,sellmann2015phase}.
In the double-moir\'e case layer plays the role of spin, gate-controlled vertical displacement fields play the 
role of magnetic field, and the physics is enriched by the long-range nature of dipolar pseudospin interactions. 
The dipole crystal states we find will give rise to plateaus in layer polarization variation
with vertical displacement fields.
We do anticipate that some long-period pseudospin crystals, that are stable at the mean-field level,
will melt to yield pseudospin liquid states, 
or possibly states with interlayer phase coherence.  
In quantum Hall bilayers, for example, charge density wave states are 
predicted by mean-field theory \cite{cote1992broken,chen1992correlated,brey1990energy} 
at intermediate layer separations, but seem to be preempted in 
reality \cite{spielman2000resonantly,kellogg2004vanishing,schliemann2001strong} by a first order transition between a small $d$ uniform density superfluid, and a large $d$ fluid state that has 
neither interlayer coherence nor crystalline order. 
The long-period states rely on dipolar interactions between widely separated neighbors, which are much weaker than nearest-neighbor interactions, so the system gains little energy by forming these states.  

Since these states have very low entropy, they can in any case appear only at very low temperatures. 
Based on the Monte Carlo simulations in Ref.~\cite{maik2012quantum}, we can estimate the melting temperature $T_m$ of the crystal states on the triangular lattice by neglecting $J^{\perp}$, which is weak.
Using a typical experimental value for the moir\'e lattice constant $a_{M}=\SI{8}{nm}$, we estimate that 
the largest ordering temperature occurs for $P=1/3$ ($2/3$ of the holes in one layer and $1/3$ in the other),
where we find that 
$T_m\approx 0.25e^2d^2/2\epsilon a_{M}^3\approx \SI{3}{K}$.  The second highest 
critical temperature occurs for $P=1/2$ ($3/4$ of the holes in one layer and $1/4$ in the other layer) 
for which $T_m\approx 0.05e^2d^2/2\epsilon a_{M}^3\approx \SI{0.5}{K}$. $T_m$ is likely reduced compared 
to these estimates by imperfections in the double moir\'e structure, for example 
misalignment of two moir\'e superlattices (see discussions below), but can be enhanced by increasing the layer separation $d$. 

In our study we have ignored the spin degree of freedom, implicitly
assuming that the energy scale of spin ordering is much lower than that of layer pseudospin ordering. 
From the point of view of strong-coupling $t^2/U$ expansions, 
this is true when the onsite repulsion $U_{\uparrow\uparrow}(0)$ (where the arrows are pseudospin labels) for holes in the same layer is much stronger than $U_{\uparrow\downarrow}(0)$, the onsite repulsion for holes in different layers. In this case we can treat spin order as a perturbative effect on top of layer pseudospin ordering. 
Since Pauli blocking occurs only for hopping processes within the same layer, the layer-polarized state is able to gain more energy than other states by suitably arranging its spins. 
Therefore when spin is taken into account, we expect that our phase diagrams for layer pseudospin ordering will stay largely unchanged, except that the layer-polarized regions in the phase diagram will expand while
the weakly polarized regions, like the LAF state region, will shrink. 
The spin ordering of the layer-polarized state has been studied in previous work on TMD moir\'es \cite{hu2021competing,morales2021non,pan2020quantum,zang2021hartree}.
The spin ordering properties of other states and its interplay \cite{zhang2021su4,zhang2022doping} with layer pseudospin 
ordering are left for future work.

In our model we have assumed that the two moir\'e layers are perfectly aligned horizontally. To our knowledge there is currently no experimental technique to control the relative alignment between two moir\'es \footnote{A closely related system is near-\SI{60}{\degree}-twisted TMD homobilayers \cite{xu2022tunable} in which interlayer tunneling is suppressed.}.  However there is recent 
evidence that moir\'e self-alignment can occur \cite{carr2020ultraheavy} 
in the course of double moir\'es device processing procedures. 
If the two moir\'e patterns are laterally displaced by a distance $s$ that is small compared to the moir\'e period, the low-energy subspace still consists of states with one hole per site, and the XXZ spin model \eqref{eq:H_XXZ} receives perturbative corrections. The $t^2/U$ terms in the coupling constants \eqref{eq:J_ij} stay unchanged, except that the value of onsite repulsion $U_{\uparrow\downarrow}(0)$ is reduced. The lateral displacement modifies the form of intersite Coulomb repulsion, and therefore pseudospin couplings. Consider two sites $i$ and $j$. The lateral displacement between two layers breaks the degeneracy between the two states $\ket{\uparrow_i \downarrow_j}$ and $\ket{\downarrow_i \uparrow_j}$ and is captured by an extra term in the Hamiltonian that is proportional to $\tau_i^z - \tau_j^z$. All such terms add up to zero at first order due to lattice symmetry. Higher-order corrections lead to anisotropic pseudospin coupling $J_{ij}^z$ and staggered sublattice potential for honeycomb lattice systems. The explicit derivation of these results is shown in Appendix~\ref{app:displaced_xxz}. Since the corrections start at second order, we expect that our results stay qualitatively unchanged for systems with a small lateral displacement between two moir\'es. On the other hand, if the lateral displacement is comparable to the moir\'e lattice constant, the localized pseudospin analogy does not work and the results will change qualitatively, see Appendix~\ref{app:displaced_xxz} and \ref{app:large_s} for more discussions.

\begin{figure}
    \centering
    \includegraphics[width=0.9\linewidth]{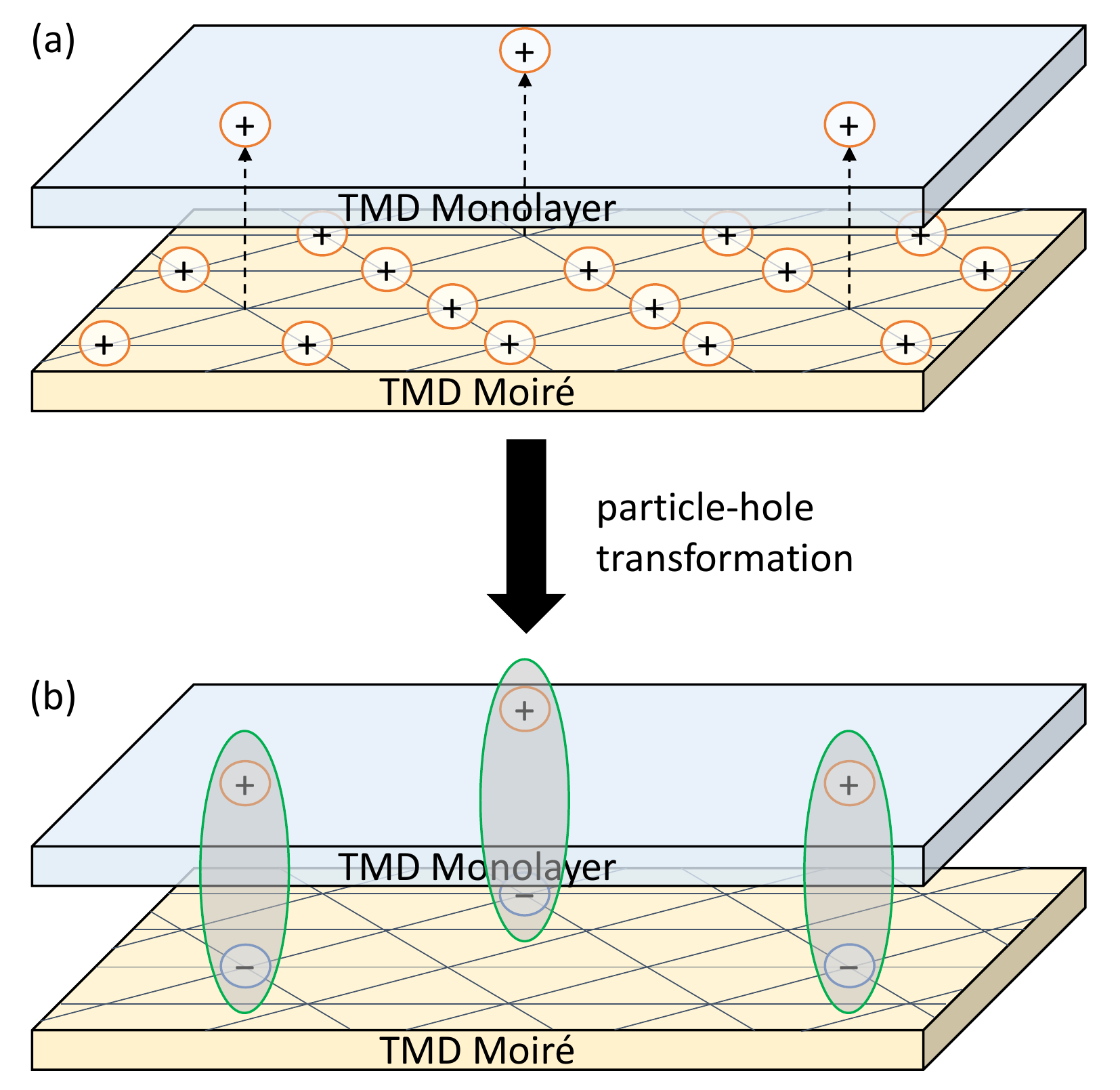}
    \caption{TMD monolayer-moir\'e coupled system at total hole filling factor $\nu=1$. (a) When the majority of doped holes are localized in the moir\'e layer forming a lattice Wigner crystal, the rest of the holes in the monolayer will stay near the region right on top of the empty sites in the moir\'e layer so that Coulomb repulsion energy is minimized. In this case the system is similar to the double-moir\'es in which the two moir\'e layers are perfectly aligned. (b) A particle-hole transformation in the moir\'e layer turns the system into a dilute gas of dipolar excitons that can be described by the hard-core boson Hubbard model.}
    \label{fig:mono_moire}
\end{figure}

\footnotetext[5]{Most physical boson systems have negative hopping parameters, corresponding to ferromagnetic in-plane coupling $J^{\perp}<0$ in the spin model. The antiferromagnetic pseudospin coupling in our system leads to frustration and richer phase diagrams that do not exist in most boson systems.}

Our study of perfectly aligned double-moir\'es also provides insight for another system which consists of a TMD moir\'e bilayer and a TMD monolayer separated by a dielectric barrier (Fig.~\ref{fig:mono_moire}). As before we consider the total filling factor such that on average one hole is present at each moir\'e site, but limit our study to the case where nearly all holes are located in the moir\'e layer. When a hole goes from the moir\'e layer to the monolayer, interaction with nearby sites limits its in-plane motion to be localized near the moir\'e site it comes from, as shown in Fig.~\ref{fig:mono_moire}(a). In other words, interaction effects produce an effective moir\'e potential in the monolayer that is perfectly aligned with that in the moir\'e layer. If we take the Mott insulator state in the moir\'e layer as the vacuum state, then the low-energy states of the system are built from the vacuum by dilute particle-hole excitations that take holes from the moir\'e layer to the monolayer. This is more clearly seen after performing a particle-hole transformation in the moir\'e layer, as shown in Fig.~\ref{fig:mono_moire}(b). The low-energy subspace of the system is the same as that of the double-moir\'e system near the layer-polarized limit, so we expect the two systems to have similar behavior in this regime. The particle-hole excitations are bosonic in nature but cannot doubly occupy a single site, so the low-energy behavior of the system is described by the hard-core boson Hubbard model \cite{yamamoto2012quantum,hebert2001quantum,schmid2002finite,jordan2009numerical,hen2010phase,Note5}. Due to the existence of an exact mapping \cite{matsubara1956lattice} between the hard-core boson Hubbard model and the XXZ spin model, the monolayer-moir\'e system can be also described by the XXZ model near the moir\'e-layer-polarized limit. Recent experiments \cite{gu2021dipolar,zhang2021correlated} have found excitonic insulator behavior of monolayer-moir\'e coupled systems near the moir\'e-layer-polarized limit, and we expect that future work based on this system can explore a larger portion of our phase diagrams.

    
\textit{Note added}: A recent experiment \cite{zeng2022doublemoire} found correlated insulating states in TMD double-moir\'{e}s formed by angle-aligned WS$_2$/bilayer WSe$_2$/WS$_2$ multilayers at hole filling factor $\nu=1$ as well as some fractional fillings.

\begin{acknowledgements}
The authors thank Jie Shan and Kin Fai Mak for helpful discussions. N.W. thanks Philipp Hauke and Maciej Lewenstein
for explaining Ref.~\onlinecite{maik2012quantum}. This work was supported by the U.S. Department of Energy, Office of Science, Basic Energy Sciences, under Award \# DE-SC0022106.
\end{acknowledgements}

\appendix

\begin{widetext}

\section{Derivation of projected Hartree-Fock equations} \label{app:hf}
We derive the projected Hartree-Fock equations starting from the microscopic Hamiltonian in Sec.~\ref{sec:model}. The single-particle physics of each TMD moir\'e layer is described by the continuum model \eqref{eq:H0}. When the moir\'e modulation potential $\Delta(\bm r)$ is sufficiently strong, the top few moir\'e valence bands are very flat and separated from lower bands by a large gap \cite{wu2018hubbard,angeli2021gamma}. The moir\'e states are related to the plane-wave states by a unitary transformation:
\begin{equation} \label{eq:moire_basis}
c_{nl}^{\dagger}(\bm k) = \sum_{\bm g} u_{n\bm g}^{(l)}(\bm k) a_{l}^{\dagger}(\bm k + \bm g),
\end{equation}
with the inverse transformation
\begin{equation}
a_{l}^{\dagger}(\bm k + \bm g) = \sum_{n} u_{n\bm g}^{(l)*}(\bm k) c_{nl}^{\dagger}(\bm k) ,
\end{equation}
Here $c^{\dagger}$ and $a^{\dagger}$ are the creation operators of moir\'e states and plane-wave states respectively, $n$ is the moir\'e band index, $l$ is the layer index, $\bm k$ labels momentum inside the moir\'e Brillouin zone, and $\bm g$ is the moir\'e reciprocal lattice vector. The $u$-coefficients are obtained by diagonalizing the continuum model Hamiltonian \eqref{eq:H0} in the plane-wave basis. In the new basis
\begin{equation}
H_0 = \sum_{nl\bm k} \epsilon_{nl}(\bm k) c_{nl}^{\dagger}(\bm k) c_{nl}(\bm k).
\end{equation}
The interaction Hamiltonian \eqref{eq:coulomb} can be written in the new basis as
\begin{equation} \label{eq:H1_c}
H_1 = \frac{1}{2A} \sum_{l' l \, \bm q} V_{l'l}(\bm q) \sum_{n'n \atop m'm} \sum_{\bm k' \bm k} \Lambda_{n'n}^{(l)}(\bm k + \bm q, \bm k) \Lambda_{m'm}^{(l')}(\bm k' - \bm q, \bm k') c_{n'l}^{\dagger}(\bm k + \bm q) c_{m'l'}^{\dagger}(\bm k' - \bm q) c_{ml'}(\bm k') c_{nl}(\bm k),
\end{equation}
where
\begin{equation}
\Lambda_{n'n}^{(l)}(\bm k',\bm k) = \sum_{\bm g} u_{n'\bm g}^{(l)*}(\bm k') u_{n\bm g}^{(l)}(\bm k).
\end{equation}
In the above equations we have extended the domains of $c^{\dagger}$ and $u$ outside the moir\'e Brillouin zone by defining
\begin{equation}
c_{nl}^{\dagger}(\bm k + \bm g) = c_{nl}^{\dagger}(\bm k), \quad u_{n\bm g'}(\bm k + \bm g) = u_{n,\bm g + \bm g'}(\bm k),
\end{equation}
so that Eq.~\eqref{eq:moire_basis} remains true for any momentum $\bm k + \bm g$. The Coulomb interaction is taken as the gate-screened form \cite{kang2020non}, with metallic gates on both sides of the double-moir\'e system separated by distance $d_g$:
\begin{align}
V_{ll'}(q) &= \frac{2\pi e^2}{\epsilon q} \frac{(e^{qd}-e^{-qd_g})(e^{-qd}-e^{-qd_g})}{1-e^{-2qd_g}}, \quad l=l', \\
V_{ll'}(q) &= \frac{2\pi e^2}{\epsilon q} \frac{e^{qd}(e^{-qd}-e^{-qd_g})^2}{1-e^{-2qd_g}}, \quad l\ne l',
\end{align}
which in the limit of $d_g\to\infty$ reduces to the unscreened form \eqref{eq:V_q}. In our numerical calculations we take $d_g = \SI{100}{nm}$.

With the Hartree-Fock approximation we decompose the interaction Hamiltonian \eqref{eq:H1_c} into the Hartree term
\begin{equation}
\Sigma_H = \frac 1A \sum_{l'l\,\bm g} \sum_{n'n\atop m'm} \sum_{\bm k' \bm k} V_{l'l}(\bm g) \Lambda_{n'n}^{(l)}(\bm k+\bm g,\bm k) \Lambda_{m'm}^{(l')}(\bm k'-\bm g,\bm k') \rho_{m'l'}^{ml'}(\bm k') c_{n'l}^{\dagger}(\bm k) c_{nl}(\bm k),
\end{equation}
and the Fock term
\begin{equation}
\Sigma_F = -\frac 1A \sum_{l'l\,\bm g} \sum_{n'n\atop m'm} \sum_{\bm k' \bm k} V_{l'l}(\bm g+\bm k'-\bm k) \Lambda_{m'n}^{(l)}(\bm k'+\bm g,\bm k) \Lambda_{n'm}^{(l')}(\bm k-\bm g,\bm k') \rho_{m'l}^{ml'}(\bm k') c_{n'l'}^{\dagger}(\bm k) c_{nl}(\bm k),
\end{equation}
with the density matrix defined as
\begin{equation}
\rho_{n'l'}^{nl}(\bm k) = \langle c_{n'l'}^{\dagger}(\bm k) c_{nl}(\bm k) \rangle,
\end{equation}
where $\langle\dots\rangle$ implies ground-state expectation values. The mean-field solutions are obtained by numerically solving the mean-field Hamiltonian $H_{\rm MF} = H_0 + \Sigma_H + \Sigma_F$ self-consistently, with band index $n$ running over only the top few isolated moir\'e bands. In our calculations we keep one band for each layer for triangular lattice systems and two for honeycomb lattice systems. Including more bands will make some quantitative changes in quantities such as the gap size, but will not change our conclusions qualitatively.

\section{Continuum model parameters} \label{app:parameters}

The single-particle physics of valence band holes is described by the continuum model Hamiltonian
\begin{equation} \label{eq:H0}
H_0 = -\frac{\hbar^2 \bm k^2}{2m} + \Delta(\bm r).
\end{equation}
The moir\'{e} potential $\Delta(\bm r)$ is given by the Fourier expansion
\begin{equation}
    \Delta(\mathbf{r})=\sum_{s=1}^{\infty} \sum_{j=1,3,5} 2V_{s} \cos \left(\mathbf{g}_{j}^{s} \cdot \mathbf{r}+\phi_{s}\right),
\end{equation}
where $\bm g_j^s$ for $j=1,2,\dots,6$ are the six moir\'e reciprocal lattice vectors in the $s$th momentum shell related by $C_6$ rotational symmetry: $\bm g_{j+1}^s=C_{6}\bm g_{j}^{s}$. $s=1,2,\dots$ labels $\bm g$-vectors with increasing magnitudes; in practice it is often a good approximation to keep only one or a few momentum shells in the Fourier expansion of moir\'e potentials. 

For the TMD homobilayer moir\'{e}, we choose the material WS$_2$ with $m=0.87m_{e},V_{1}=\SI{33.5}{meV},V_{2}=\SI{4.0}{meV},V_{3}=\SI{5.5}{meV},V_{s>3}=0,\phi_{s}=\SI{180}{\degree}$ and the lattice constant $a_{0}=$\SI{3.18}{\AA} obtained from \textit{ab initio} calculations \cite{angeli2021gamma}.

For the heterobilayer, doped holes populate the valence band of only one of the layers (the active layer). The other layer generates a potential with the moir\'{e} periodicity that affects holes in the active layer. We assume the active layer is WSe$_{2}$ with $m=0.35m_{e}$. We neglect the lattice mismatch between two layers and use the lattice constant of WSe$_{2}$ $a_{0}=\SI{3.32}{\AA}$. Including a small lattice mismatch in our calculation will change the relation between the twist angle $\theta$ and the moir\'{e} lattice constant $a_{M}$, but the phase diagram for given $a_{M}$ should not change. We take the following moir\'{e} potential parameters \cite{morales2021non}: $V_{1}=\SI{11}{meV},V_{s>1}=0,\phi_1=-\SI{94}{\degree}$. The strength of the modulation potential depends on the material choice of the inactive layer and can be effectively modified by pressure. 

\section{Spatial distribution of layer pseudospins} \label{app:spin_plot}

\begin{figure}
   \includegraphics[width=0.9\columnwidth]{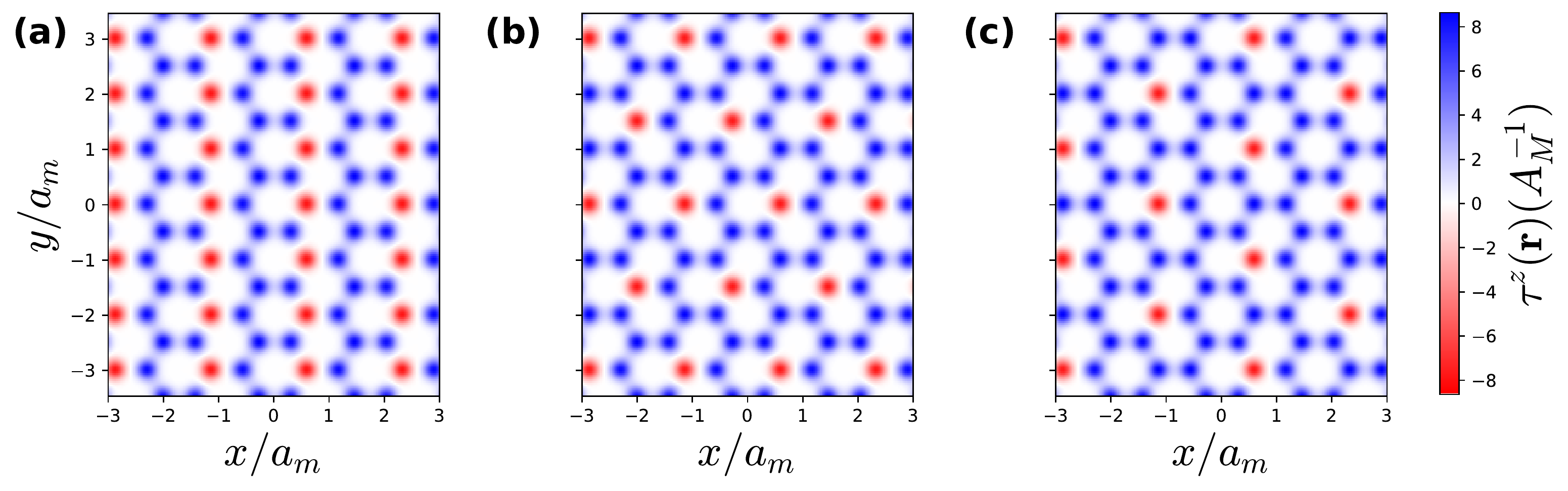}
    \caption{The spatial distribution of layer pseudospins for several dipole crystal states on double AA-stacked WS$_2$ homobilayer moir\'{e}s. Each moir\'e has twist angle $\theta=\SI{2.5}{\degree}$. The blue and red regions are holes in the top ($\uparrow$) and bottom ($\downarrow$) layers, respectively. The layer polarization $P=(n_{\uparrow}-n_{\downarrow})/(n_{\uparrow}+n_{\downarrow})$ for the three plots are: (a) $P=1/2$; (b) $P=2/3$; (c) $P=3/4$.
}
    \label{fig:spin_G}
\end{figure}
\begin{figure}
   \includegraphics[width=0.9\columnwidth]{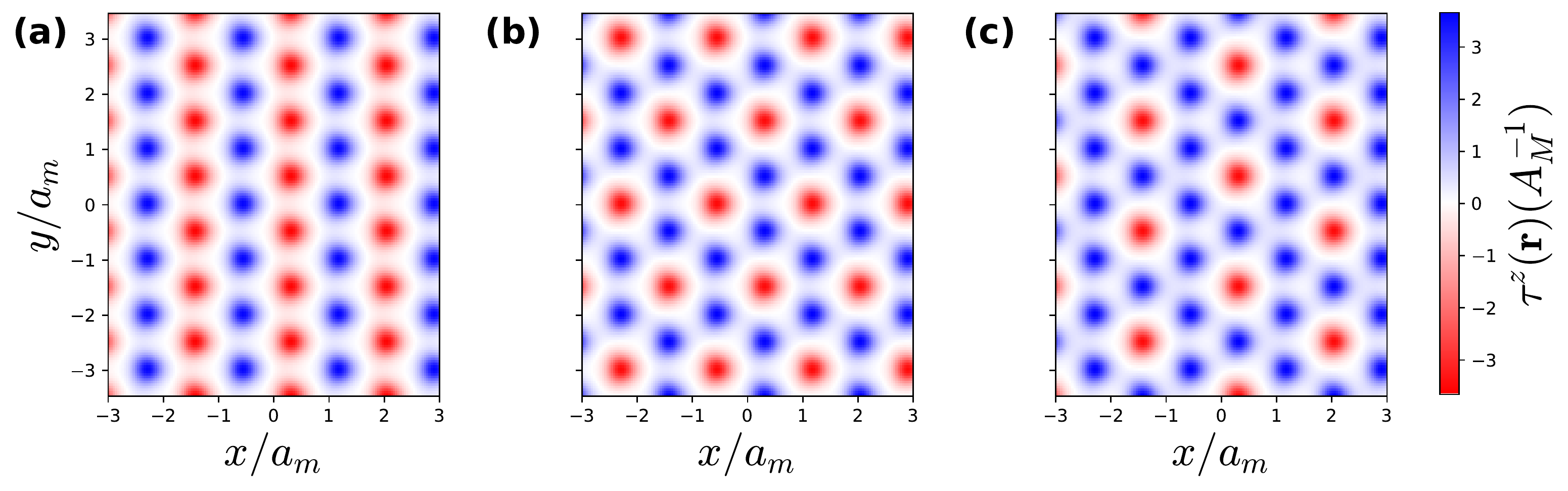}
    \caption{The spatial distribution of layer pseudospins for several dipole crystal states on double heterobilayer moir\'{e}s. Each moir\'e has twist angle $\theta=\SI{2.5}{\degree}$. The layer polarization (a) $P=0$; (b) $P=1/3$; (c) $P=1/2$.
}
    \label{fig:spin_K}
\end{figure}

In this appendix, we plot the spatial distribution of layer polarization of the dipole crystal states. The local layer polarization $\tau^{z}(\bm r)=\langle \Psi^{\dagger}(\bm r)\tau^z\Psi(\bm r) \rangle$ is computed as follows:
\begin{equation}
    \tau^{z}(\bm r)= \sum_{n\bm g,n'\bm g',\bm k}\left(\rho_{n'\uparrow}^{n\uparrow}(\bm k)-\rho_{n'\downarrow}^{n\downarrow}(\bm k)\right)u_{n'\bm g'}(\bm k)u_{n\bm g}^{*}(\bm k)e^{i(\bm g-\bm g')\cdot\bm r}
\end{equation}
where we defined the electron annilation operator $\Psi(\bm r)$ and expanded it in the band basis $c_{nl}(\bm k)$. The layer superscripts of the Bloch wave functions $u$ are dropped because we only study the case when two layers are identical in the main text.
For sufficiently large $V_z$, the holes are polarized in the top layer ($\tau^z(\bm r)>0$). As $V_z$ decreases, part of holes are transferred to the bottom layer. Our mean-field calculations show that a large density of holes in the minority layer could crystallize and become localized in red regions, as plotted in Figs.~\ref{fig:spin_G} and \ref{fig:spin_K}. Fig.~\ref{fig:spin_G} shows the spatial distribution of layer pseudospins for several dipole crystal states on double WS$_2$ homobilayer moir\'{e}s, while Fig.~\ref{fig:spin_K} is for double heterobilayer moir\'{e}s. 

In the following, let us focus on the triangular lattice case. Fig.~\ref{fig:spin_K}(b) is a three-sublattice phase with $\sqrt{3}\times\sqrt{3}$ unit cells ($\uparrow\uparrow\downarrow$ state), which is the most robust crystal state in the phase diagram (Fig.~\ref{fig:K}(b)) in the main text. The other robust crystal state ($\uparrow\uparrow\uparrow\downarrow$) is shown in Fig.~\ref{fig:spin_K}(c), while the two-sublattice state at small displacement field in Fig.~\ref{fig:spin_K}(a) consists alternating stripes with opposite layer polarizations.

Note that these crystal states are obtained in $\sqrt{3}\times\sqrt{3}$ and $2\times 2$ supercells. There certainly exists more crystal states with the same total layer polarization but longer periodicity, which are not captured in our calculations.

\section{Modified XXZ model with lateral displacement} \label{app:displaced_xxz}
In this appendix we derive the effective spin model that describes the double-moir\'e system with a lateral displacement. As long as the displacement is small compared to the moir\'e period such that the onsite repulsion $U_{\uparrow\downarrow}(0)$ is much larger than the other relevant energy scales, it remains a good approximation to project the Hamiltonian onto the low-energy subspace with one hole per site and expand in powers of $t^2/U$. The derivation of the $t^2/U$-expansion is standard and leads to the same result as in the perfectly-aligned double-moir\'es, so in the following we focus on the modification of the Coulomb contributions to the pseudospin couplings parameters.
 
Assume the top moir\'e is shifted by an in-plane vector $\bm s$ relative to the bottom moir\'e. Consider two site $i$ and $j$. While the states $\ket{\uparrow_i \uparrow_j}$ and $\ket{\downarrow_i \downarrow_j}$ remain degenerate ($U_{ij,\uparrow\uparrow} = U_{ij,\downarrow\downarrow}$), the degeneracy between states $\ket{\uparrow_i \downarrow_j}$ and $\ket{\downarrow_i \uparrow_j}$ is broken ($U_{ij,\uparrow\downarrow} \ne U_{ij,\downarrow\uparrow}$). This results in an extra term in the XXZ Hamiltonian \eqref{eq:H_XXZ}:
\begin{equation} \label{eq:dH_XXZ}
\Delta\mathcal{H}_{\rm XXZ} = \sum_{i<j} \frac{U_{ij,\uparrow\downarrow} - U_{ij,\downarrow\uparrow}}{4} (\tau_i^z - \tau_j^z) = \sum_{i} \frac{B_i^z}{2} \tau_i^z.
\end{equation}
The final form of the above equation shows that the lateral displacement produces an effective magnetic field $B_i^z$ for the pseudospin at site $i$. Since $\tau_i^z$ and $-\tau_j^z$ always come in pairs in the summands, the average of $B_i^z$ over all lattice sites must be zero. For triangular lattice systems, since all lattice sites are equivalent, $B_i^z=0$ identically for all $i$. For honeycomb lattice systems, in contrast, $B_i^z$ can take nonzero and opposite values on the two sublattices and act as a staggered field. The Coulomb part of the pseudospin coupling parameter is also changed by the lateral displacement and takes the form
\begin{equation} \label{eq:Jz_new}
J_{ij}^z \big|_{\rm Coul} = (2U_{ij,\uparrow\uparrow} - U_{ij,\uparrow\downarrow} - U_{ij,\downarrow\uparrow})/4.
\end{equation}

To get explicit expressions we approximate the repulsion energy by the simple Coulomb forms
\begin{align}
U_{ij,\uparrow\uparrow} &= U_{ij,\downarrow\downarrow} \approx e^2/\epsilon R_{ij}, \\
U_{ij,\uparrow\downarrow} &\approx e^2/\epsilon \sqrt{(\bm R_{ij} + \bm s)^2 + d^2}, \\
U_{ij,\downarrow\uparrow} &\approx e^2/\epsilon \sqrt{(\bm R_{ij} - \bm s)^2 + d^2},
\end{align}
where $\bm R_{ij} = \bm R_i - \bm R_j$. We then expand $U_{ij,\uparrow\downarrow}$ and $U_{ij,\downarrow\uparrow}$ in powers of $\bm s$:
\begin{align}
U_{ij,\uparrow\downarrow} &\approx \frac{e^2}{\epsilon\sqrt{R_{ij}^2+d^2}} \left[1 - \frac{2\bm R_{ij}\cdot\bm s + s^2}{2(R_{ij}^2+d^2)} + \frac 32\, \frac{(\bm R_{ij}\cdot\bm s)^2 + (\bm R_{ij}\cdot\bm s)s^2}{(R_{ij}^2+d^2)^2} - \frac 52 \frac{(\bm R_{ij}\cdot\bm s)^3}{(R_{ij}^2+d^2)^3} + O(s^4) \right], \\
U_{ij,\downarrow\uparrow} &\approx \frac{e^2}{\epsilon\sqrt{R_{ij}^2+d^2}} \left[1 + \frac{2\bm R_{ij}\cdot\bm s - s^2}{2(R_{ij}^2+d^2)} + \frac 32\, \frac{(\bm R_{ij}\cdot\bm s)^2 - (\bm R_{ij}\cdot\bm s)s^2}{(R_{ij}^2+d^2)^2} + \frac 52 \frac{(\bm R_{ij}\cdot\bm s)^3}{(R_{ij}^2+d^2)^3} + O(s^4) \right].
\end{align}
Plugging into Eq.~\eqref{eq:Jz_new}, we get the correction to the pseudospin coupling parameter
\begin{equation} \label{eq:deltaJ}
\Delta J_{ij}^z \approx \frac{e^2}{4\epsilon (R_{ij}^2+d^2)^{3/2}} \left[s^2 - \frac{3(\bm R_{ij}\cdot\bm s)^2}{R_{ij}^2+d^2} + O(s^4) \right]
\end{equation}
compared to the perfectly aligned double-moir\'es. Notice that the second term in the square bracket produces an anisotropic contribution to $J_{ij}^z$. From Eq.~\eqref{eq:dH_XXZ} we get the effective magnetic field
\begin{equation} \label{eq:Bz}
B_i^z = \sum_{j\ne i} \frac{U_{ij,\uparrow\downarrow} - U_{ij,\downarrow\uparrow}}{2} \approx -\frac 52 \sum_{j\ne i} \frac{e^2}{\epsilon (R_{ij}^2+d^2)^{7/2}} (\bm R_{ij} \cdot\bm s)^3 + O(s^5).
\end{equation}
In getting the final form we have made use of the identity $\sum_{j\ne i} f(R_{ij})\bm R_{ij} = 0$ which holds for any function $f$ for both triangular and honeycomb lattices. It is straightforward to show that Eq.~\eqref{eq:Bz} vanishes to all orders in $\bm s$ for triangular lattice systems, but produces a staggered field on the two sublattices in honeycomb lattices. However, since all corrections only start at second order, we expect that a small shift $\bm s$ does not make qualitative differences to our results.  

\section{An example of largely misaligned double-moir\'{e}s} \label{app:large_s}
When the lateral shift $s$ is comparable with the moir\'e lattice constant $a_M$, our XXZ model analysis is no longer applicable and the results become very different. As an example, Fig.~\ref{fig:K_shift} shows the phase diagram of a triangular lattice system with lateral shift $\bm s = (\sqrt{3}/4,0)a_{M}$ (we choose our coordinate system such that the moir\'e lattice vectors point along $(\sqrt{3}/2,\pm 1/2)$).
We find that at large twist angles the phase diagram is similar to that of the perfectly aligned double-moir\'{e}s (Fig.~\ref{fig:K}), indicating that moir\'e misalignment is less important in the weak modulation limit. 
However, the large lateral shift $s$ does make some notable differences in part of the phase diagram.

\begin{figure}
   \includegraphics[width=0.3\columnwidth]{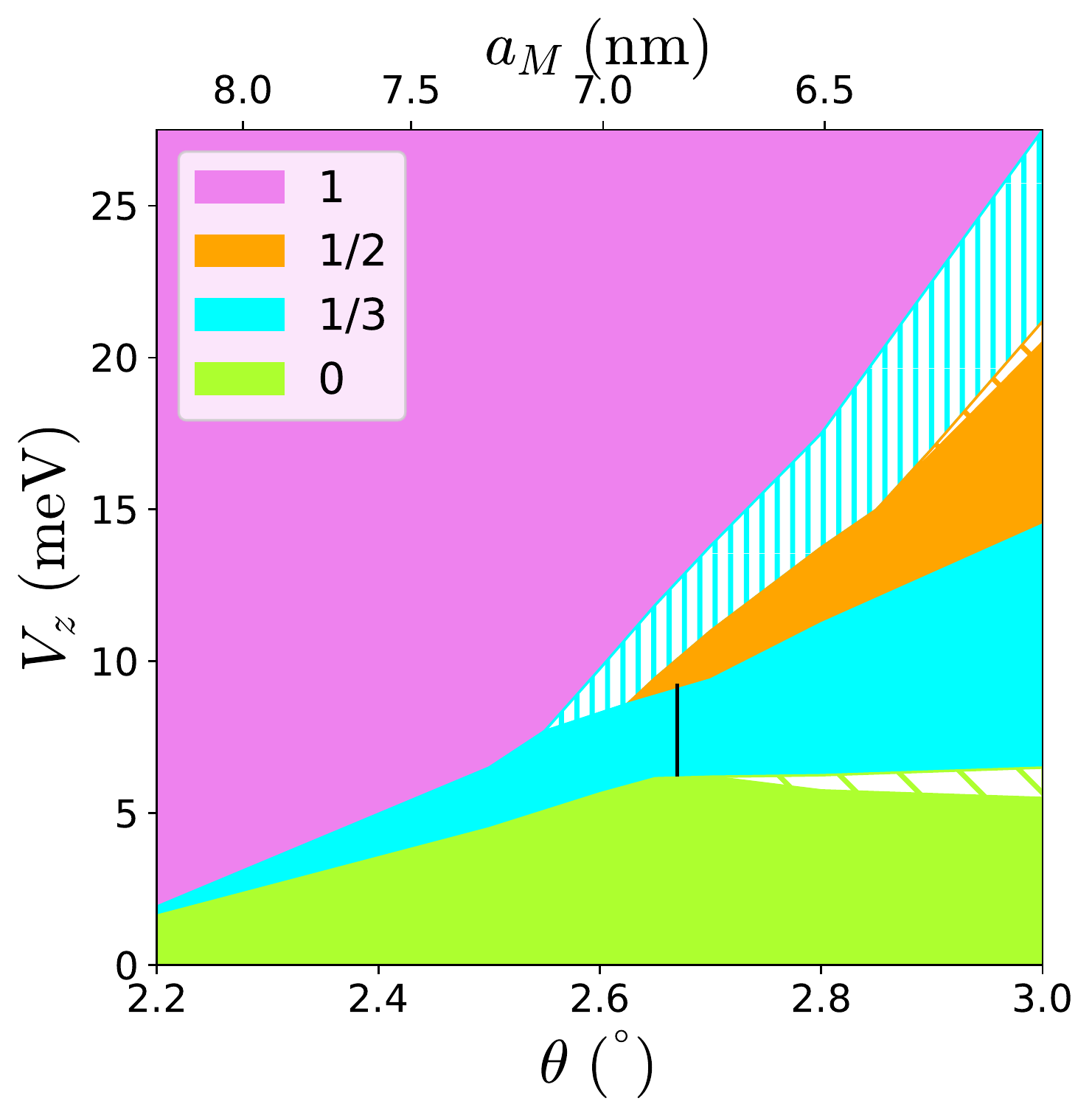}
    \caption{The phase diagram of triangular-lattice double-moir\'{e}s at filling factor $\nu=1$ in the plane of displacement field $V_z$ and twist angle $\theta$. Two moir\'{e}s are laterally shifted by $\bm s = (\sqrt{3}/4,0)a_{M}$ and vertically separated by $d=\SI{2}{nm}$. The solid filled regions represent states with no interlayer coherence and different colors represent different layer polarizations $P=(n_{\uparrow}-n_{\downarrow})/(n_{\uparrow}+n_{\downarrow})$. The hatched regions represent supersolid states that break both layer-$U(1)$ and translational symmetries. At large twist angles with small $s/d$, the phase diagram maintains all phases of the perfectly aligned double-moir\'{e}s, except that the layer-coherent supersolid state between layer-incoherent $P=1/3$ crystal and $P=0$ stripe states is now replaced by a layer-coherent stripe state in the hatched green region. 
    However, at small twist angles with large $s/d$, the $P=1/3$ stripe state has lower energy than the $\sqrt{3}\times\sqrt{3}$ crystal state. These two layer-incoherent states, both with $P=1/3$, are separated by a vertical solid black line. 
}
    \label{fig:K_shift}
\end{figure}

First, a lateral shift reduces the threshold displacement field for the layer-polarized state. This can be understood from electrostatic considerations. Distributing charges into both layers minimizes Coulomb repulsion for perfectly aligned systems, but when $s$ is large such configurations would increase Coulomb repulsion due to the inhomogeneous in-plane charge distribution. Therefore the layer-polarized state is more favorable at large $s/d$.

Second, the layer-coherent states disappear for large $a_{M}$ (small $\theta$) in Fig.~\ref{fig:K_shift}. Note that this result does not contradict our previous arguments for the existence of layer-coherent states near the layer-polarized state in the perfect alignment case, which rely on the hard-core boson Hubbard model analogy. This analogy breaks down in the presence of a large lateral shift due to the existence of two or three types of near-degenerate electron-hole excitations (with different in-plane dipole moments) whose interactions can be either repulsive or attractive. In such a system electron-hole excitations prefer to form collectively and the layer polarization has a sudden jump at the boundary of the layer-polarized state, as shown in the phase diagram for $\theta<\SI{2.5}{\degree}$.

Third, we find that stripe states become more favorable compared to dipole crystal states. At $\theta\lesssim\SI{2.6}{\degree}$, the dipole crystals completely disappear. A lateral shift between two moir\'{e}s generically breaks the $C_3$ rotation symmetry about the $z$-axis and suppresses the $C_{3}$-symmetric dipole crystal states in Fig.~\ref{fig:spin_K}(b)(c). In contrast, stripe states are not $C_3$-symmetric and are less susceptible to the rotation symmetry breaking induced by the lateral shift. Since dipole crystals become energetically less favorable, their critical temperatures are expected to decrease with the lateral shift between two moir\'{e}s.

\end{widetext}

\bibliography{reference.bib}

\begin{thebibliography}{56}%
\makeatletter
\providecommand \@ifxundefined [1]{%
 \@ifx{#1\undefined}
}%
\providecommand \@ifnum [1]{%
 \ifnum #1\expandafter \@firstoftwo
 \else \expandafter \@secondoftwo
 \fi
}%
\providecommand \@ifx [1]{%
 \ifx #1\expandafter \@firstoftwo
 \else \expandafter \@secondoftwo
 \fi
}%
\providecommand \natexlab [1]{#1}%
\providecommand \enquote  [1]{``#1''}%
\providecommand \bibnamefont  [1]{#1}%
\providecommand \bibfnamefont [1]{#1}%
\providecommand \citenamefont [1]{#1}%
\providecommand \href@noop [0]{\@secondoftwo}%
\providecommand \href [0]{\begingroup \@sanitize@url \@href}%
\providecommand \@href[1]{\@@startlink{#1}\@@href}%
\providecommand \@@href[1]{\endgroup#1\@@endlink}%
\providecommand \@sanitize@url [0]{\catcode `\\12\catcode `\$12\catcode
  `\&12\catcode `\#12\catcode `\^12\catcode `\_12\catcode `\%12\relax}%
\providecommand \@@startlink[1]{}%
\providecommand \@@endlink[0]{}%
\providecommand \url  [0]{\begingroup\@sanitize@url \@url }%
\providecommand \@url [1]{\endgroup\@href {#1}{\urlprefix }}%
\providecommand \urlprefix  [0]{URL }%
\providecommand \Eprint [0]{\href }%
\providecommand \doibase [0]{https://doi.org/}%
\providecommand \selectlanguage [0]{\@gobble}%
\providecommand \bibinfo  [0]{\@secondoftwo}%
\providecommand \bibfield  [0]{\@secondoftwo}%
\providecommand \translation [1]{[#1]}%
\providecommand \BibitemOpen [0]{}%
\providecommand \bibitemStop [0]{}%
\providecommand \bibitemNoStop [0]{.\EOS\space}%
\providecommand \EOS [0]{\spacefactor3000\relax}%
\providecommand \BibitemShut  [1]{\csname bibitem#1\endcsname}%
\let\auto@bib@innerbib\@empty
\bibitem [{\citenamefont {Eisenstein}\ and\ \citenamefont
  {MacDonald}(2004)}]{eisenstein2004bose}%
  \BibitemOpen
  \bibfield  {author} {\bibinfo {author} {\bibfnamefont {J.}~\bibnamefont
  {Eisenstein}}\ and\ \bibinfo {author} {\bibfnamefont {A.}~\bibnamefont
  {MacDonald}},\ }\bibfield  {title} {\bibinfo {title} {{Bose-Einstein
  condensation of excitons in bilayer electron systems}},\ }\href@noop {}
  {\bibfield  {journal} {\bibinfo  {journal} {Nature}\ }\textbf {\bibinfo
  {volume} {432}},\ \bibinfo {pages} {691} (\bibinfo {year}
  {2004})}\BibitemShut {NoStop}%
\bibitem [{\citenamefont {Eisenstein}(2014)}]{eisenstein2014exciton}%
  \BibitemOpen
  \bibfield  {author} {\bibinfo {author} {\bibfnamefont {J.}~\bibnamefont
  {Eisenstein}},\ }\bibfield  {title} {\bibinfo {title} {{Exciton condensation
  in bilayer quantum Hall systems}},\ }\href@noop {} {\bibfield  {journal}
  {\bibinfo  {journal} {Annu. Rev. Condens. Matter Phys.}\ }\textbf {\bibinfo
  {volume} {5}},\ \bibinfo {pages} {159} (\bibinfo {year} {2014})}\BibitemShut
  {NoStop}%
\bibitem [{\citenamefont {Cao}\ \emph {et~al.}(2018{\natexlab{a}})\citenamefont
  {Cao}, \citenamefont {Fatemi}, \citenamefont {Demir}, \citenamefont {Fang},
  \citenamefont {Tomarken}, \citenamefont {Luo}, \citenamefont
  {Sanchez-Yamagishi}, \citenamefont {Watanabe}, \citenamefont {Taniguchi},
  \citenamefont {Kaxiras} \emph {et~al.}}]{cao2018correlated}%
  \BibitemOpen
  \bibfield  {author} {\bibinfo {author} {\bibfnamefont {Y.}~\bibnamefont
  {Cao}}, \bibinfo {author} {\bibfnamefont {V.}~\bibnamefont {Fatemi}},
  \bibinfo {author} {\bibfnamefont {A.}~\bibnamefont {Demir}}, \bibinfo
  {author} {\bibfnamefont {S.}~\bibnamefont {Fang}}, \bibinfo {author}
  {\bibfnamefont {S.~L.}\ \bibnamefont {Tomarken}}, \bibinfo {author}
  {\bibfnamefont {J.~Y.}\ \bibnamefont {Luo}}, \bibinfo {author} {\bibfnamefont
  {J.~D.}\ \bibnamefont {Sanchez-Yamagishi}}, \bibinfo {author} {\bibfnamefont
  {K.}~\bibnamefont {Watanabe}}, \bibinfo {author} {\bibfnamefont
  {T.}~\bibnamefont {Taniguchi}}, \bibinfo {author} {\bibfnamefont
  {E.}~\bibnamefont {Kaxiras}}, \emph {et~al.},\ }\bibfield  {title} {\bibinfo
  {title} {Correlated insulator behaviour at half-filling in magic-angle
  graphene superlattices},\ }\href@noop {} {\bibfield  {journal} {\bibinfo
  {journal} {Nature}\ }\textbf {\bibinfo {volume} {556}},\ \bibinfo {pages}
  {80} (\bibinfo {year} {2018}{\natexlab{a}})}\BibitemShut {NoStop}%
\bibitem [{\citenamefont {Cao}\ \emph {et~al.}(2018{\natexlab{b}})\citenamefont
  {Cao}, \citenamefont {Fatemi}, \citenamefont {Fang}, \citenamefont
  {Watanabe}, \citenamefont {Taniguchi}, \citenamefont {Kaxiras},\ and\
  \citenamefont {Jarillo-Herrero}}]{cao2018unconventional}%
  \BibitemOpen
  \bibfield  {author} {\bibinfo {author} {\bibfnamefont {Y.}~\bibnamefont
  {Cao}}, \bibinfo {author} {\bibfnamefont {V.}~\bibnamefont {Fatemi}},
  \bibinfo {author} {\bibfnamefont {S.}~\bibnamefont {Fang}}, \bibinfo {author}
  {\bibfnamefont {K.}~\bibnamefont {Watanabe}}, \bibinfo {author}
  {\bibfnamefont {T.}~\bibnamefont {Taniguchi}}, \bibinfo {author}
  {\bibfnamefont {E.}~\bibnamefont {Kaxiras}},\ and\ \bibinfo {author}
  {\bibfnamefont {P.}~\bibnamefont {Jarillo-Herrero}},\ }\bibfield  {title}
  {\bibinfo {title} {Unconventional superconductivity in magic-angle graphene
  superlattices},\ }\href@noop {} {\bibfield  {journal} {\bibinfo  {journal}
  {Nature}\ }\textbf {\bibinfo {volume} {556}},\ \bibinfo {pages} {43}
  (\bibinfo {year} {2018}{\natexlab{b}})}\BibitemShut {NoStop}%
\bibitem [{\citenamefont {Regan}\ \emph {et~al.}(2020)\citenamefont {Regan},
  \citenamefont {Wang}, \citenamefont {Jin}, \citenamefont {Utama},
  \citenamefont {Gao}, \citenamefont {Wei}, \citenamefont {Zhao}, \citenamefont
  {Zhao}, \citenamefont {Zhang}, \citenamefont {Yumigeta} \emph
  {et~al.}}]{regan2020mott}%
  \BibitemOpen
  \bibfield  {author} {\bibinfo {author} {\bibfnamefont {E.~C.}\ \bibnamefont
  {Regan}}, \bibinfo {author} {\bibfnamefont {D.}~\bibnamefont {Wang}},
  \bibinfo {author} {\bibfnamefont {C.}~\bibnamefont {Jin}}, \bibinfo {author}
  {\bibfnamefont {M.~I.~B.}\ \bibnamefont {Utama}}, \bibinfo {author}
  {\bibfnamefont {B.}~\bibnamefont {Gao}}, \bibinfo {author} {\bibfnamefont
  {X.}~\bibnamefont {Wei}}, \bibinfo {author} {\bibfnamefont {S.}~\bibnamefont
  {Zhao}}, \bibinfo {author} {\bibfnamefont {W.}~\bibnamefont {Zhao}}, \bibinfo
  {author} {\bibfnamefont {Z.}~\bibnamefont {Zhang}}, \bibinfo {author}
  {\bibfnamefont {K.}~\bibnamefont {Yumigeta}}, \emph {et~al.},\ }\bibfield
  {title} {\bibinfo {title} {Mott and generalized {Wigner crystal states in
  WSe$_2$/WS$_2$ moir{\'e} superlattices}},\ }\href@noop {} {\bibfield
  {journal} {\bibinfo  {journal} {Nature}\ }\textbf {\bibinfo {volume} {579}},\
  \bibinfo {pages} {359} (\bibinfo {year} {2020})}\BibitemShut {NoStop}%
\bibitem [{\citenamefont {Tang}\ \emph {et~al.}(2020)\citenamefont {Tang},
  \citenamefont {Li}, \citenamefont {Li}, \citenamefont {Xu}, \citenamefont
  {Liu}, \citenamefont {Barmak}, \citenamefont {Watanabe}, \citenamefont
  {Taniguchi}, \citenamefont {MacDonald}, \citenamefont {Shan} \emph
  {et~al.}}]{tang2020simulation}%
  \BibitemOpen
  \bibfield  {author} {\bibinfo {author} {\bibfnamefont {Y.}~\bibnamefont
  {Tang}}, \bibinfo {author} {\bibfnamefont {L.}~\bibnamefont {Li}}, \bibinfo
  {author} {\bibfnamefont {T.}~\bibnamefont {Li}}, \bibinfo {author}
  {\bibfnamefont {Y.}~\bibnamefont {Xu}}, \bibinfo {author} {\bibfnamefont
  {S.}~\bibnamefont {Liu}}, \bibinfo {author} {\bibfnamefont {K.}~\bibnamefont
  {Barmak}}, \bibinfo {author} {\bibfnamefont {K.}~\bibnamefont {Watanabe}},
  \bibinfo {author} {\bibfnamefont {T.}~\bibnamefont {Taniguchi}}, \bibinfo
  {author} {\bibfnamefont {A.~H.}\ \bibnamefont {MacDonald}}, \bibinfo {author}
  {\bibfnamefont {J.}~\bibnamefont {Shan}}, \emph {et~al.},\ }\bibfield
  {title} {\bibinfo {title} {Simulation of hubbard model physics in wse2/ws2
  moir{\'e} superlattices},\ }\href@noop {} {\bibfield  {journal} {\bibinfo
  {journal} {Nature}\ }\textbf {\bibinfo {volume} {579}},\ \bibinfo {pages}
  {353} (\bibinfo {year} {2020})}\BibitemShut {NoStop}%
\bibitem [{\citenamefont {Xu}\ \emph {et~al.}(2020)\citenamefont {Xu},
  \citenamefont {Liu}, \citenamefont {Rhodes}, \citenamefont {Watanabe},
  \citenamefont {Taniguchi}, \citenamefont {Hone}, \citenamefont {Elser},
  \citenamefont {Mak},\ and\ \citenamefont {Shan}}]{xu2020correlated}%
  \BibitemOpen
  \bibfield  {author} {\bibinfo {author} {\bibfnamefont {Y.}~\bibnamefont
  {Xu}}, \bibinfo {author} {\bibfnamefont {S.}~\bibnamefont {Liu}}, \bibinfo
  {author} {\bibfnamefont {D.~A.}\ \bibnamefont {Rhodes}}, \bibinfo {author}
  {\bibfnamefont {K.}~\bibnamefont {Watanabe}}, \bibinfo {author}
  {\bibfnamefont {T.}~\bibnamefont {Taniguchi}}, \bibinfo {author}
  {\bibfnamefont {J.}~\bibnamefont {Hone}}, \bibinfo {author} {\bibfnamefont
  {V.}~\bibnamefont {Elser}}, \bibinfo {author} {\bibfnamefont {K.~F.}\
  \bibnamefont {Mak}},\ and\ \bibinfo {author} {\bibfnamefont {J.}~\bibnamefont
  {Shan}},\ }\bibfield  {title} {\bibinfo {title} {Correlated insulating states
  at fractional fillings of moir{\'e} superlattices},\ }\href@noop {}
  {\bibfield  {journal} {\bibinfo  {journal} {Nature}\ }\textbf {\bibinfo
  {volume} {587}},\ \bibinfo {pages} {214} (\bibinfo {year}
  {2020})}\BibitemShut {NoStop}%
\bibitem [{\citenamefont {Spielman}\ \emph {et~al.}(2000)\citenamefont
  {Spielman}, \citenamefont {Eisenstein}, \citenamefont {Pfeiffer},\ and\
  \citenamefont {West}}]{spielman2000resonantly}%
  \BibitemOpen
  \bibfield  {author} {\bibinfo {author} {\bibfnamefont {I.~B.}\ \bibnamefont
  {Spielman}}, \bibinfo {author} {\bibfnamefont {J.~P.}\ \bibnamefont
  {Eisenstein}}, \bibinfo {author} {\bibfnamefont {L.~N.}\ \bibnamefont
  {Pfeiffer}},\ and\ \bibinfo {author} {\bibfnamefont {K.~W.}\ \bibnamefont
  {West}},\ }\bibfield  {title} {\bibinfo {title} {Resonantly enhanced
  tunneling in a double layer quantum hall ferromagnet},\ }\href
  {https://doi.org/10.1103/PhysRevLett.84.5808} {\bibfield  {journal} {\bibinfo
   {journal} {Phys. Rev. Lett.}\ }\textbf {\bibinfo {volume} {84}},\ \bibinfo
  {pages} {5808} (\bibinfo {year} {2000})}\BibitemShut {NoStop}%
\bibitem [{\citenamefont {Kellogg}\ \emph {et~al.}(2004)\citenamefont
  {Kellogg}, \citenamefont {Eisenstein}, \citenamefont {Pfeiffer},\ and\
  \citenamefont {West}}]{kellogg2004vanishing}%
  \BibitemOpen
  \bibfield  {author} {\bibinfo {author} {\bibfnamefont {M.}~\bibnamefont
  {Kellogg}}, \bibinfo {author} {\bibfnamefont {J.~P.}\ \bibnamefont
  {Eisenstein}}, \bibinfo {author} {\bibfnamefont {L.~N.}\ \bibnamefont
  {Pfeiffer}},\ and\ \bibinfo {author} {\bibfnamefont {K.~W.}\ \bibnamefont
  {West}},\ }\bibfield  {title} {\bibinfo {title} {Vanishing hall resistance at
  high magnetic field in a double-layer two-dimensional electron system},\
  }\href {https://doi.org/10.1103/PhysRevLett.93.036801} {\bibfield  {journal}
  {\bibinfo  {journal} {Phys. Rev. Lett.}\ }\textbf {\bibinfo {volume} {93}},\
  \bibinfo {pages} {036801} (\bibinfo {year} {2004})}\BibitemShut {NoStop}%
\bibitem [{\citenamefont {Tutuc}\ \emph {et~al.}(2004)\citenamefont {Tutuc},
  \citenamefont {Shayegan},\ and\ \citenamefont {Huse}}]{tutuc2004counterflow}%
  \BibitemOpen
  \bibfield  {author} {\bibinfo {author} {\bibfnamefont {E.}~\bibnamefont
  {Tutuc}}, \bibinfo {author} {\bibfnamefont {M.}~\bibnamefont {Shayegan}},\
  and\ \bibinfo {author} {\bibfnamefont {D.~A.}\ \bibnamefont {Huse}},\
  }\bibfield  {title} {\bibinfo {title} {{Counterflow Measurements in Strongly
  Correlated GaAs Hole Bilayers: Evidence for Electron-Hole Pairing}},\ }\href
  {https://doi.org/10.1103/PhysRevLett.93.036802} {\bibfield  {journal}
  {\bibinfo  {journal} {Phys. Rev. Lett.}\ }\textbf {\bibinfo {volume} {93}},\
  \bibinfo {pages} {036802} (\bibinfo {year} {2004})}\BibitemShut {NoStop}%
\bibitem [{\citenamefont {Bistritzer}\ and\ \citenamefont
  {MacDonald}(2011)}]{bistritzer2011moire}%
  \BibitemOpen
  \bibfield  {author} {\bibinfo {author} {\bibfnamefont {R.}~\bibnamefont
  {Bistritzer}}\ and\ \bibinfo {author} {\bibfnamefont {A.~H.}\ \bibnamefont
  {MacDonald}},\ }\bibfield  {title} {\bibinfo {title} {Moir{\'e} bands in
  twisted double-layer graphene},\ }\href@noop {} {\bibfield  {journal}
  {\bibinfo  {journal} {Proceedings of the National Academy of Sciences}\
  }\textbf {\bibinfo {volume} {108}},\ \bibinfo {pages} {12233} (\bibinfo
  {year} {2011})}\BibitemShut {NoStop}%
\bibitem [{\citenamefont {Wu}\ \emph {et~al.}(2018)\citenamefont {Wu},
  \citenamefont {Lovorn}, \citenamefont {Tutuc},\ and\ \citenamefont
  {MacDonald}}]{wu2018hubbard}%
  \BibitemOpen
  \bibfield  {author} {\bibinfo {author} {\bibfnamefont {F.}~\bibnamefont
  {Wu}}, \bibinfo {author} {\bibfnamefont {T.}~\bibnamefont {Lovorn}}, \bibinfo
  {author} {\bibfnamefont {E.}~\bibnamefont {Tutuc}},\ and\ \bibinfo {author}
  {\bibfnamefont {A.~H.}\ \bibnamefont {MacDonald}},\ }\bibfield  {title}
  {\bibinfo {title} {Hubbard model physics in transition metal dichalcogenide
  moir\'e bands},\ }\href {https://doi.org/10.1103/PhysRevLett.121.026402}
  {\bibfield  {journal} {\bibinfo  {journal} {Phys. Rev. Lett.}\ }\textbf
  {\bibinfo {volume} {121}},\ \bibinfo {pages} {026402} (\bibinfo {year}
  {2018})}\BibitemShut {NoStop}%
\bibitem [{\citenamefont {Angeli}\ and\ \citenamefont
  {MacDonald}(2021)}]{angeli2021gamma}%
  \BibitemOpen
  \bibfield  {author} {\bibinfo {author} {\bibfnamefont {M.}~\bibnamefont
  {Angeli}}\ and\ \bibinfo {author} {\bibfnamefont {A.~H.}\ \bibnamefont
  {MacDonald}},\ }\bibfield  {title} {\bibinfo {title} {{$\Gamma$ valley
  transition metal dichalcogenide moir{\'e} bands}},\ }\href@noop {} {\bibfield
   {journal} {\bibinfo  {journal} {Proceedings of the National Academy of
  Sciences}\ }\textbf {\bibinfo {volume} {118}} (\bibinfo {year}
  {2021})}\BibitemShut {NoStop}%
\bibitem [{\citenamefont {Xian}\ \emph {et~al.}(2021)\citenamefont {Xian},
  \citenamefont {Claassen}, \citenamefont {Kiese}, \citenamefont {Scherer},
  \citenamefont {Trebst}, \citenamefont {Kennes},\ and\ \citenamefont
  {Rubio}}]{xian2021realization}%
  \BibitemOpen
  \bibfield  {author} {\bibinfo {author} {\bibfnamefont {L.}~\bibnamefont
  {Xian}}, \bibinfo {author} {\bibfnamefont {M.}~\bibnamefont {Claassen}},
  \bibinfo {author} {\bibfnamefont {D.}~\bibnamefont {Kiese}}, \bibinfo
  {author} {\bibfnamefont {M.~M.}\ \bibnamefont {Scherer}}, \bibinfo {author}
  {\bibfnamefont {S.}~\bibnamefont {Trebst}}, \bibinfo {author} {\bibfnamefont
  {D.~M.}\ \bibnamefont {Kennes}},\ and\ \bibinfo {author} {\bibfnamefont
  {A.}~\bibnamefont {Rubio}},\ }\bibfield  {title} {\bibinfo {title}
  {Realization of nearly dispersionless bands with strong orbital anisotropy
  from destructive interference in twisted bilayer mos2},\ }\href@noop {}
  {\bibfield  {journal} {\bibinfo  {journal} {Nature communications}\ }\textbf
  {\bibinfo {volume} {12}},\ \bibinfo {pages} {1} (\bibinfo {year}
  {2021})}\BibitemShut {NoStop}%
\bibitem [{\citenamefont {Kune{\v{s}}}(2015)}]{kunevs2015excitonic}%
  \BibitemOpen
  \bibfield  {author} {\bibinfo {author} {\bibfnamefont {J.}~\bibnamefont
  {Kune{\v{s}}}},\ }\bibfield  {title} {\bibinfo {title} {Excitonic
  condensation in systems of strongly correlated electrons},\ }\href@noop {}
  {\bibfield  {journal} {\bibinfo  {journal} {Journal of Physics: Condensed
  Matter}\ }\textbf {\bibinfo {volume} {27}},\ \bibinfo {pages} {333201}
  (\bibinfo {year} {2015})}\BibitemShut {NoStop}%
\bibitem [{\citenamefont {Batista}(2002)}]{batista2002electronic}%
  \BibitemOpen
  \bibfield  {author} {\bibinfo {author} {\bibfnamefont {C.~D.}\ \bibnamefont
  {Batista}},\ }\bibfield  {title} {\bibinfo {title} {Electronic
  ferroelectricity in the {Falicov-Kimball} model},\ }\href
  {https://doi.org/10.1103/PhysRevLett.89.166403} {\bibfield  {journal}
  {\bibinfo  {journal} {Phys. Rev. Lett.}\ }\textbf {\bibinfo {volume} {89}},\
  \bibinfo {pages} {166403} (\bibinfo {year} {2002})}\BibitemShut {NoStop}%
\bibitem [{\citenamefont {Portengen}\ \emph {et~al.}(1996)\citenamefont
  {Portengen}, \citenamefont {\"Ostreich},\ and\ \citenamefont
  {Sham}}]{portengen1996theory}%
  \BibitemOpen
  \bibfield  {author} {\bibinfo {author} {\bibfnamefont {T.}~\bibnamefont
  {Portengen}}, \bibinfo {author} {\bibfnamefont {T.}~\bibnamefont
  {\"Ostreich}},\ and\ \bibinfo {author} {\bibfnamefont {L.~J.}\ \bibnamefont
  {Sham}},\ }\bibfield  {title} {\bibinfo {title} {Theory of electronic
  ferroelectricity},\ }\href {https://doi.org/10.1103/PhysRevB.54.17452}
  {\bibfield  {journal} {\bibinfo  {journal} {Phys. Rev. B}\ }\textbf {\bibinfo
  {volume} {54}},\ \bibinfo {pages} {17452} (\bibinfo {year}
  {1996})}\BibitemShut {NoStop}%
\bibitem [{\citenamefont {Batista}\ \emph {et~al.}(2004)\citenamefont
  {Batista}, \citenamefont {Gubernatis}, \citenamefont
  {Bon\ifmmode~\check{c}\else \v{c}\fi{}a},\ and\ \citenamefont
  {Lin}}]{batista2004intermediate}%
  \BibitemOpen
  \bibfield  {author} {\bibinfo {author} {\bibfnamefont {C.~D.}\ \bibnamefont
  {Batista}}, \bibinfo {author} {\bibfnamefont {J.~E.}\ \bibnamefont
  {Gubernatis}}, \bibinfo {author} {\bibfnamefont {J.}~\bibnamefont
  {Bon\ifmmode~\check{c}\else \v{c}\fi{}a}},\ and\ \bibinfo {author}
  {\bibfnamefont {H.~Q.}\ \bibnamefont {Lin}},\ }\bibfield  {title} {\bibinfo
  {title} {Intermediate coupling theory of electronic ferroelectricity},\
  }\href {https://doi.org/10.1103/PhysRevLett.92.187601} {\bibfield  {journal}
  {\bibinfo  {journal} {Phys. Rev. Lett.}\ }\textbf {\bibinfo {volume} {92}},\
  \bibinfo {pages} {187601} (\bibinfo {year} {2004})}\BibitemShut {NoStop}%
\bibitem [{\citenamefont {Farka\ifmmode~\check{s}\else
  \v{s}\fi{}ovsk\'y}(2008)}]{farkasovsky2008hartree}%
  \BibitemOpen
  \bibfield  {author} {\bibinfo {author} {\bibfnamefont {P.}~\bibnamefont
  {Farka\ifmmode~\check{s}\else \v{s}\fi{}ovsk\'y}},\ }\bibfield  {title}
  {\bibinfo {title} {Hartree-fock study of electronic ferroelectricity in the
  falicov-kimball model with $f\text{\ensuremath{-}}f$ hopping},\ }\href
  {https://doi.org/10.1103/PhysRevB.77.155130} {\bibfield  {journal} {\bibinfo
  {journal} {Phys. Rev. B}\ }\textbf {\bibinfo {volume} {77}},\ \bibinfo
  {pages} {155130} (\bibinfo {year} {2008})}\BibitemShut {NoStop}%
\bibitem [{\citenamefont {Kaneko}\ \emph {et~al.}(2013)\citenamefont {Kaneko},
  \citenamefont {Ejima}, \citenamefont {Fehske},\ and\ \citenamefont
  {Ohta}}]{kaneko2013exact}%
  \BibitemOpen
  \bibfield  {author} {\bibinfo {author} {\bibfnamefont {T.}~\bibnamefont
  {Kaneko}}, \bibinfo {author} {\bibfnamefont {S.}~\bibnamefont {Ejima}},
  \bibinfo {author} {\bibfnamefont {H.}~\bibnamefont {Fehske}},\ and\ \bibinfo
  {author} {\bibfnamefont {Y.}~\bibnamefont {Ohta}},\ }\bibfield  {title}
  {\bibinfo {title} {Exact-diagonalization study of exciton condensation in
  electron bilayers},\ }\href {https://doi.org/10.1103/PhysRevB.88.035312}
  {\bibfield  {journal} {\bibinfo  {journal} {Phys. Rev. B}\ }\textbf {\bibinfo
  {volume} {88}},\ \bibinfo {pages} {035312} (\bibinfo {year}
  {2013})}\BibitemShut {NoStop}%
\bibitem [{\citenamefont {Holtschneider}\ and\ \citenamefont
  {Selke}(2007)}]{holtschneider2007biconical}%
  \BibitemOpen
  \bibfield  {author} {\bibinfo {author} {\bibfnamefont {M.}~\bibnamefont
  {Holtschneider}}\ and\ \bibinfo {author} {\bibfnamefont {W.}~\bibnamefont
  {Selke}},\ }\bibfield  {title} {\bibinfo {title} {Biconical structures in
  two-dimensional anisotropic heisenberg antiferromagnets},\ }\href
  {https://doi.org/10.1103/PhysRevB.76.220405} {\bibfield  {journal} {\bibinfo
  {journal} {Phys. Rev. B}\ }\textbf {\bibinfo {volume} {76}},\ \bibinfo
  {pages} {220405} (\bibinfo {year} {2007})}\BibitemShut {NoStop}%
\bibitem [{\citenamefont {Yamashita}(1972)}]{yamashita1972field}%
  \BibitemOpen
  \bibfield  {author} {\bibinfo {author} {\bibfnamefont {N.}~\bibnamefont
  {Yamashita}},\ }\bibfield  {title} {\bibinfo {title} {Field induced phase
  transitions in uniaxial antiferromagnets},\ }\href@noop {} {\bibfield
  {journal} {\bibinfo  {journal} {Journal of the Physical Society of Japan}\
  }\textbf {\bibinfo {volume} {32}},\ \bibinfo {pages} {610} (\bibinfo {year}
  {1972})}\BibitemShut {NoStop}%
\bibitem [{\citenamefont {Matsuda}\ and\ \citenamefont
  {Tsuneto}(1970)}]{matsuda1970off}%
  \BibitemOpen
  \bibfield  {author} {\bibinfo {author} {\bibfnamefont {H.}~\bibnamefont
  {Matsuda}}\ and\ \bibinfo {author} {\bibfnamefont {T.}~\bibnamefont
  {Tsuneto}},\ }\bibfield  {title} {\bibinfo {title} {Off-diagonal long-range
  order in solids},\ }\href@noop {} {\bibfield  {journal} {\bibinfo  {journal}
  {Progress of Theoretical Physics Supplement}\ }\textbf {\bibinfo {volume}
  {46}},\ \bibinfo {pages} {411} (\bibinfo {year} {1970})}\BibitemShut
  {NoStop}%
\bibitem [{\citenamefont {Liu}\ and\ \citenamefont
  {Fisher}(1973)}]{liu1973quantum}%
  \BibitemOpen
  \bibfield  {author} {\bibinfo {author} {\bibfnamefont {K.-S.}\ \bibnamefont
  {Liu}}\ and\ \bibinfo {author} {\bibfnamefont {M.~E.}\ \bibnamefont
  {Fisher}},\ }\bibfield  {title} {\bibinfo {title} {Quantum lattice gas and
  the existence of a supersolid},\ }\href@noop {} {\bibfield  {journal}
  {\bibinfo  {journal} {Journal of Low Temperature Physics}\ }\textbf {\bibinfo
  {volume} {10}},\ \bibinfo {pages} {655} (\bibinfo {year} {1973})}\BibitemShut
  {NoStop}%
\bibitem [{\citenamefont {Bruce}\ and\ \citenamefont
  {Aharony}(1975)}]{bruce1975coupled}%
  \BibitemOpen
  \bibfield  {author} {\bibinfo {author} {\bibfnamefont {A.~D.}\ \bibnamefont
  {Bruce}}\ and\ \bibinfo {author} {\bibfnamefont {A.}~\bibnamefont
  {Aharony}},\ }\bibfield  {title} {\bibinfo {title} {Coupled order parameters,
  symmetry-breaking irrelevant scaling fields, and tetracritical points},\
  }\href {https://doi.org/10.1103/PhysRevB.11.478} {\bibfield  {journal}
  {\bibinfo  {journal} {Phys. Rev. B}\ }\textbf {\bibinfo {volume} {11}},\
  \bibinfo {pages} {478} (\bibinfo {year} {1975})}\BibitemShut {NoStop}%
\bibitem [{\citenamefont {Yamamoto}\ \emph {et~al.}(2012)\citenamefont
  {Yamamoto}, \citenamefont {Masaki},\ and\ \citenamefont
  {Danshita}}]{yamamoto2012quantum}%
  \BibitemOpen
  \bibfield  {author} {\bibinfo {author} {\bibfnamefont {D.}~\bibnamefont
  {Yamamoto}}, \bibinfo {author} {\bibfnamefont {A.}~\bibnamefont {Masaki}},\
  and\ \bibinfo {author} {\bibfnamefont {I.}~\bibnamefont {Danshita}},\
  }\bibfield  {title} {\bibinfo {title} {Quantum phases of hardcore bosons with
  long-range interactions on a square lattice},\ }\href
  {https://doi.org/10.1103/PhysRevB.86.054516} {\bibfield  {journal} {\bibinfo
  {journal} {Phys. Rev. B}\ }\textbf {\bibinfo {volume} {86}},\ \bibinfo
  {pages} {054516} (\bibinfo {year} {2012})}\BibitemShut {NoStop}%
\bibitem [{\citenamefont {Capogrosso-Sansone}\ \emph
  {et~al.}(2010)\citenamefont {Capogrosso-Sansone}, \citenamefont {Trefzger},
  \citenamefont {Lewenstein}, \citenamefont {Zoller},\ and\ \citenamefont
  {Pupillo}}]{capogrosso2010optical}%
  \BibitemOpen
  \bibfield  {author} {\bibinfo {author} {\bibfnamefont {B.}~\bibnamefont
  {Capogrosso-Sansone}}, \bibinfo {author} {\bibfnamefont {C.}~\bibnamefont
  {Trefzger}}, \bibinfo {author} {\bibfnamefont {M.}~\bibnamefont
  {Lewenstein}}, \bibinfo {author} {\bibfnamefont {P.}~\bibnamefont {Zoller}},\
  and\ \bibinfo {author} {\bibfnamefont {G.}~\bibnamefont {Pupillo}},\
  }\bibfield  {title} {\bibinfo {title} {Quantum phases of cold polar molecules
  in 2d optical lattices},\ }\href
  {https://doi.org/10.1103/PhysRevLett.104.125301} {\bibfield  {journal}
  {\bibinfo  {journal} {Phys. Rev. Lett.}\ }\textbf {\bibinfo {volume} {104}},\
  \bibinfo {pages} {125301} (\bibinfo {year} {2010})}\BibitemShut {NoStop}%
\bibitem [{\citenamefont {Hubbard}(1978)}]{hubbard1978wigner}%
  \BibitemOpen
  \bibfield  {author} {\bibinfo {author} {\bibfnamefont {J.}~\bibnamefont
  {Hubbard}},\ }\bibfield  {title} {\bibinfo {title} {Generalized wigner
  lattices in one dimension and some applications to tetracyanoquinodimethane
  (tcnq) salts},\ }\href {https://doi.org/10.1103/PhysRevB.17.494} {\bibfield
  {journal} {\bibinfo  {journal} {Phys. Rev. B}\ }\textbf {\bibinfo {volume}
  {17}},\ \bibinfo {pages} {494} (\bibinfo {year} {1978})}\BibitemShut
  {NoStop}%
\bibitem [{\citenamefont {Fisher}\ and\ \citenamefont
  {Selke}(1980)}]{fisher1980infinitely}%
  \BibitemOpen
  \bibfield  {author} {\bibinfo {author} {\bibfnamefont {M.~E.}\ \bibnamefont
  {Fisher}}\ and\ \bibinfo {author} {\bibfnamefont {W.}~\bibnamefont {Selke}},\
  }\bibfield  {title} {\bibinfo {title} {Infinitely many commensurate phases in
  a simple ising model},\ }\href {https://doi.org/10.1103/PhysRevLett.44.1502}
  {\bibfield  {journal} {\bibinfo  {journal} {Phys. Rev. Lett.}\ }\textbf
  {\bibinfo {volume} {44}},\ \bibinfo {pages} {1502} (\bibinfo {year}
  {1980})}\BibitemShut {NoStop}%
\bibitem [{\citenamefont {Bak}\ and\ \citenamefont
  {Bruinsma}(1982)}]{bak1982ising}%
  \BibitemOpen
  \bibfield  {author} {\bibinfo {author} {\bibfnamefont {P.}~\bibnamefont
  {Bak}}\ and\ \bibinfo {author} {\bibfnamefont {R.}~\bibnamefont {Bruinsma}},\
  }\bibfield  {title} {\bibinfo {title} {One-dimensional ising model and the
  complete devil's staircase},\ }\href
  {https://doi.org/10.1103/PhysRevLett.49.249} {\bibfield  {journal} {\bibinfo
  {journal} {Phys. Rev. Lett.}\ }\textbf {\bibinfo {volume} {49}},\ \bibinfo
  {pages} {249} (\bibinfo {year} {1982})}\BibitemShut {NoStop}%
\bibitem [{\citenamefont {Morales-Dur{\'a}n}\ \emph {et~al.}(2021)\citenamefont
  {Morales-Dur{\'a}n}, \citenamefont {Hu}, \citenamefont {Potasz},\ and\
  \citenamefont {MacDonald}}]{morales2021non}%
  \BibitemOpen
  \bibfield  {author} {\bibinfo {author} {\bibfnamefont {N.}~\bibnamefont
  {Morales-Dur{\'a}n}}, \bibinfo {author} {\bibfnamefont {N.~C.}\ \bibnamefont
  {Hu}}, \bibinfo {author} {\bibfnamefont {P.}~\bibnamefont {Potasz}},\ and\
  \bibinfo {author} {\bibfnamefont {A.~H.}\ \bibnamefont {MacDonald}},\
  }\bibfield  {title} {\bibinfo {title} {Non-local interactions in moir\'e
  {Hubbard} systems},\ }\href@noop {} {\bibfield  {journal} {\bibinfo
  {journal} {arXiv preprint arXiv:2108.03313}\ } (\bibinfo {year}
  {2021})}\BibitemShut {NoStop}%
\bibitem [{\citenamefont {Yamamoto}\ \emph {et~al.}(2014)\citenamefont
  {Yamamoto}, \citenamefont {Marmorini},\ and\ \citenamefont
  {Danshita}}]{yamamoto2014quantum}%
  \BibitemOpen
  \bibfield  {author} {\bibinfo {author} {\bibfnamefont {D.}~\bibnamefont
  {Yamamoto}}, \bibinfo {author} {\bibfnamefont {G.}~\bibnamefont
  {Marmorini}},\ and\ \bibinfo {author} {\bibfnamefont {I.}~\bibnamefont
  {Danshita}},\ }\bibfield  {title} {\bibinfo {title} {Quantum phase diagram of
  the triangular-lattice {$XXZ$} model in a magnetic field},\ }\href
  {https://doi.org/10.1103/PhysRevLett.112.127203} {\bibfield  {journal}
  {\bibinfo  {journal} {Phys. Rev. Lett.}\ }\textbf {\bibinfo {volume} {112}},\
  \bibinfo {pages} {127203} (\bibinfo {year} {2014})}\BibitemShut {NoStop}%
\bibitem [{\citenamefont {Sellmann}\ \emph {et~al.}(2015)\citenamefont
  {Sellmann}, \citenamefont {Zhang},\ and\ \citenamefont
  {Eggert}}]{sellmann2015phase}%
  \BibitemOpen
  \bibfield  {author} {\bibinfo {author} {\bibfnamefont {D.}~\bibnamefont
  {Sellmann}}, \bibinfo {author} {\bibfnamefont {X.-F.}\ \bibnamefont
  {Zhang}},\ and\ \bibinfo {author} {\bibfnamefont {S.}~\bibnamefont
  {Eggert}},\ }\bibfield  {title} {\bibinfo {title} {Phase diagram of the
  antiferromagnetic xxz model on the triangular lattice},\ }\href
  {https://doi.org/10.1103/PhysRevB.91.081104} {\bibfield  {journal} {\bibinfo
  {journal} {Phys. Rev. B}\ }\textbf {\bibinfo {volume} {91}},\ \bibinfo
  {pages} {081104} (\bibinfo {year} {2015})}\BibitemShut {NoStop}%
\bibitem [{\citenamefont {C\^ot\'e}\ \emph {et~al.}(1992)\citenamefont
  {C\^ot\'e}, \citenamefont {Brey},\ and\ \citenamefont
  {MacDonald}}]{cote1992broken}%
  \BibitemOpen
  \bibfield  {author} {\bibinfo {author} {\bibfnamefont {R.}~\bibnamefont
  {C\^ot\'e}}, \bibinfo {author} {\bibfnamefont {L.}~\bibnamefont {Brey}},\
  and\ \bibinfo {author} {\bibfnamefont {A.~H.}\ \bibnamefont {MacDonald}},\
  }\bibfield  {title} {\bibinfo {title} {Broken-symmetry ground states for the
  two-dimensional electron gas in a double-quantum-well system},\ }\href
  {https://doi.org/10.1103/PhysRevB.46.10239} {\bibfield  {journal} {\bibinfo
  {journal} {Phys. Rev. B}\ }\textbf {\bibinfo {volume} {46}},\ \bibinfo
  {pages} {10239} (\bibinfo {year} {1992})}\BibitemShut {NoStop}%
\bibitem [{\citenamefont {Chen}\ and\ \citenamefont
  {Quinn}(1992)}]{chen1992correlated}%
  \BibitemOpen
  \bibfield  {author} {\bibinfo {author} {\bibfnamefont {X.~M.}\ \bibnamefont
  {Chen}}\ and\ \bibinfo {author} {\bibfnamefont {J.~J.}\ \bibnamefont
  {Quinn}},\ }\bibfield  {title} {\bibinfo {title} {Correlated
  charge-density-wave states of double-quantum-well systems in a strong
  magnetic field},\ }\href {https://doi.org/10.1103/PhysRevB.45.11054}
  {\bibfield  {journal} {\bibinfo  {journal} {Phys. Rev. B}\ }\textbf {\bibinfo
  {volume} {45}},\ \bibinfo {pages} {11054} (\bibinfo {year}
  {1992})}\BibitemShut {NoStop}%
\bibitem [{\citenamefont {Brey}(1990)}]{brey1990energy}%
  \BibitemOpen
  \bibfield  {author} {\bibinfo {author} {\bibfnamefont {L.}~\bibnamefont
  {Brey}},\ }\bibfield  {title} {\bibinfo {title} {Energy spectrum and
  charge-density-wave instability of a double quantum well in a magnetic
  field},\ }\href {https://doi.org/10.1103/PhysRevLett.65.903} {\bibfield
  {journal} {\bibinfo  {journal} {Phys. Rev. Lett.}\ }\textbf {\bibinfo
  {volume} {65}},\ \bibinfo {pages} {903} (\bibinfo {year} {1990})}\BibitemShut
  {NoStop}%
\bibitem [{\citenamefont {Schliemann}\ \emph {et~al.}(2001)\citenamefont
  {Schliemann}, \citenamefont {Girvin},\ and\ \citenamefont
  {MacDonald}}]{schliemann2001strong}%
  \BibitemOpen
  \bibfield  {author} {\bibinfo {author} {\bibfnamefont {J.}~\bibnamefont
  {Schliemann}}, \bibinfo {author} {\bibfnamefont {S.~M.}\ \bibnamefont
  {Girvin}},\ and\ \bibinfo {author} {\bibfnamefont {A.~H.}\ \bibnamefont
  {MacDonald}},\ }\bibfield  {title} {\bibinfo {title} {Strong correlation to
  weak correlation phase transition in bilayer quantum hall systems},\ }\href
  {https://doi.org/10.1103/PhysRevLett.86.1849} {\bibfield  {journal} {\bibinfo
   {journal} {Phys. Rev. Lett.}\ }\textbf {\bibinfo {volume} {86}},\ \bibinfo
  {pages} {1849} (\bibinfo {year} {2001})}\BibitemShut {NoStop}%
\bibitem [{\citenamefont {Maik}\ \emph {et~al.}(2012)\citenamefont {Maik},
  \citenamefont {Hauke}, \citenamefont {Dutta}, \citenamefont {Zakrzewski},\
  and\ \citenamefont {Lewenstein}}]{maik2012quantum}%
  \BibitemOpen
  \bibfield  {author} {\bibinfo {author} {\bibfnamefont {M.}~\bibnamefont
  {Maik}}, \bibinfo {author} {\bibfnamefont {P.}~\bibnamefont {Hauke}},
  \bibinfo {author} {\bibfnamefont {O.}~\bibnamefont {Dutta}}, \bibinfo
  {author} {\bibfnamefont {J.}~\bibnamefont {Zakrzewski}},\ and\ \bibinfo
  {author} {\bibfnamefont {M.}~\bibnamefont {Lewenstein}},\ }\bibfield  {title}
  {\bibinfo {title} {Quantum spin models with long-range interactions and
  tunnelings: a quantum monte carlo study},\ }\href@noop {} {\bibfield
  {journal} {\bibinfo  {journal} {New Journal of Physics}\ }\textbf {\bibinfo
  {volume} {14}},\ \bibinfo {pages} {113006} (\bibinfo {year}
  {2012})}\BibitemShut {NoStop}%
\bibitem [{\citenamefont {Hu}\ and\ \citenamefont
  {MacDonald}(2021)}]{hu2021competing}%
  \BibitemOpen
  \bibfield  {author} {\bibinfo {author} {\bibfnamefont {N.~C.}\ \bibnamefont
  {Hu}}\ and\ \bibinfo {author} {\bibfnamefont {A.~H.}\ \bibnamefont
  {MacDonald}},\ }\bibfield  {title} {\bibinfo {title} {Competing magnetic
  states in transition metal dichalcogenide moir\'e materials},\ }\href
  {https://doi.org/10.1103/PhysRevB.104.214403} {\bibfield  {journal} {\bibinfo
   {journal} {Phys. Rev. B}\ }\textbf {\bibinfo {volume} {104}},\ \bibinfo
  {pages} {214403} (\bibinfo {year} {2021})}\BibitemShut {NoStop}%
\bibitem [{\citenamefont {Pan}\ \emph {et~al.}(2020)\citenamefont {Pan},
  \citenamefont {Wu},\ and\ \citenamefont {Das~Sarma}}]{pan2020quantum}%
  \BibitemOpen
  \bibfield  {author} {\bibinfo {author} {\bibfnamefont {H.}~\bibnamefont
  {Pan}}, \bibinfo {author} {\bibfnamefont {F.}~\bibnamefont {Wu}},\ and\
  \bibinfo {author} {\bibfnamefont {S.}~\bibnamefont {Das~Sarma}},\ }\bibfield
  {title} {\bibinfo {title} {Quantum phase diagram of a {Moir\'e-Hubbard}
  model},\ }\href {https://doi.org/10.1103/PhysRevB.102.201104} {\bibfield
  {journal} {\bibinfo  {journal} {Phys. Rev. B}\ }\textbf {\bibinfo {volume}
  {102}},\ \bibinfo {pages} {201104} (\bibinfo {year} {2020})}\BibitemShut
  {NoStop}%
\bibitem [{\citenamefont {Zang}\ \emph {et~al.}(2021)\citenamefont {Zang},
  \citenamefont {Wang}, \citenamefont {Cano},\ and\ \citenamefont
  {Millis}}]{zang2021hartree}%
  \BibitemOpen
  \bibfield  {author} {\bibinfo {author} {\bibfnamefont {J.}~\bibnamefont
  {Zang}}, \bibinfo {author} {\bibfnamefont {J.}~\bibnamefont {Wang}}, \bibinfo
  {author} {\bibfnamefont {J.}~\bibnamefont {Cano}},\ and\ \bibinfo {author}
  {\bibfnamefont {A.~J.}\ \bibnamefont {Millis}},\ }\bibfield  {title}
  {\bibinfo {title} {{Hartree-Fock Study of the Moir\'e Hubbard Model for
  Twisted Bilayer Transition Metal Dichalcogenides}},\ }\href@noop {}
  {\bibfield  {journal} {\bibinfo  {journal} {arXiv preprint arXiv:2105.11883}\
  } (\bibinfo {year} {2021})}\BibitemShut {NoStop}%
\bibitem [{\citenamefont {Zhang}\ \emph
  {et~al.}(2021{\natexlab{a}})\citenamefont {Zhang}, \citenamefont {Sheng},\
  and\ \citenamefont {Vishwanath}}]{zhang2021su4}%
  \BibitemOpen
  \bibfield  {author} {\bibinfo {author} {\bibfnamefont {Y.-H.}\ \bibnamefont
  {Zhang}}, \bibinfo {author} {\bibfnamefont {D.~N.}\ \bibnamefont {Sheng}},\
  and\ \bibinfo {author} {\bibfnamefont {A.}~\bibnamefont {Vishwanath}},\
  }\bibfield  {title} {\bibinfo {title} {{$SU(4)$ Chiral Spin Liquid, Exciton
  Supersolid, and Electric Detection in Moir\'e Bilayers}},\ }\href
  {https://doi.org/10.1103/PhysRevLett.127.247701} {\bibfield  {journal}
  {\bibinfo  {journal} {Phys. Rev. Lett.}\ }\textbf {\bibinfo {volume} {127}},\
  \bibinfo {pages} {247701} (\bibinfo {year} {2021}{\natexlab{a}})}\BibitemShut
  {NoStop}%
\bibitem [{\citenamefont {Zhang}(2022)}]{zhang2022doping}%
  \BibitemOpen
  \bibfield  {author} {\bibinfo {author} {\bibfnamefont {Y.-H.}\ \bibnamefont
  {Zhang}},\ }\bibfield  {title} {\bibinfo {title} {{Doping a Mott insulator
  with excitons in moir\'e bilayer: fractional superfluid, neutral Fermi
  surface and Mott transition}},\ }\href@noop {} {\bibfield  {journal}
  {\bibinfo  {journal} {arXiv preprint arXiv:2204.10937}\ } (\bibinfo {year}
  {2022})}\BibitemShut {NoStop}%
\bibitem [{Note1()}]{Note1}%
  \BibitemOpen
  \bibinfo {note} {A closely related system is near-\SI {60}{\degree }-twisted
  TMD homobilayers \cite {xu2022tunable} in which interlayer tunneling is
  suppressed.}\BibitemShut {Stop}%
\bibitem [{\citenamefont {Carr}\ \emph {et~al.}(2020)\citenamefont {Carr},
  \citenamefont {Li}, \citenamefont {Zhu}, \citenamefont {Kaxiras},
  \citenamefont {Sachdev},\ and\ \citenamefont
  {Kruchkov}}]{carr2020ultraheavy}%
  \BibitemOpen
  \bibfield  {author} {\bibinfo {author} {\bibfnamefont {S.}~\bibnamefont
  {Carr}}, \bibinfo {author} {\bibfnamefont {C.}~\bibnamefont {Li}}, \bibinfo
  {author} {\bibfnamefont {Z.}~\bibnamefont {Zhu}}, \bibinfo {author}
  {\bibfnamefont {E.}~\bibnamefont {Kaxiras}}, \bibinfo {author} {\bibfnamefont
  {S.}~\bibnamefont {Sachdev}},\ and\ \bibinfo {author} {\bibfnamefont
  {A.}~\bibnamefont {Kruchkov}},\ }\bibfield  {title} {\bibinfo {title}
  {Ultraheavy and ultrarelativistic dirac quasiparticles in sandwiched
  graphenes},\ }\href@noop {} {\bibfield  {journal} {\bibinfo  {journal} {Nano
  letters}\ }\textbf {\bibinfo {volume} {20}},\ \bibinfo {pages} {3030}
  (\bibinfo {year} {2020})}\BibitemShut {NoStop}%
\bibitem [{\citenamefont {H\'ebert}\ \emph {et~al.}(2001)\citenamefont
  {H\'ebert}, \citenamefont {Batrouni}, \citenamefont {Scalettar},
  \citenamefont {Schmid}, \citenamefont {Troyer},\ and\ \citenamefont
  {Dorneich}}]{hebert2001quantum}%
  \BibitemOpen
  \bibfield  {author} {\bibinfo {author} {\bibfnamefont {F.}~\bibnamefont
  {H\'ebert}}, \bibinfo {author} {\bibfnamefont {G.~G.}\ \bibnamefont
  {Batrouni}}, \bibinfo {author} {\bibfnamefont {R.~T.}\ \bibnamefont
  {Scalettar}}, \bibinfo {author} {\bibfnamefont {G.}~\bibnamefont {Schmid}},
  \bibinfo {author} {\bibfnamefont {M.}~\bibnamefont {Troyer}},\ and\ \bibinfo
  {author} {\bibfnamefont {A.}~\bibnamefont {Dorneich}},\ }\bibfield  {title}
  {\bibinfo {title} {Quantum phase transitions in the two-dimensional hardcore
  boson model},\ }\href {https://doi.org/10.1103/PhysRevB.65.014513} {\bibfield
   {journal} {\bibinfo  {journal} {Phys. Rev. B}\ }\textbf {\bibinfo {volume}
  {65}},\ \bibinfo {pages} {014513} (\bibinfo {year} {2001})}\BibitemShut
  {NoStop}%
\bibitem [{\citenamefont {Schmid}\ \emph {et~al.}(2002)\citenamefont {Schmid},
  \citenamefont {Todo}, \citenamefont {Troyer},\ and\ \citenamefont
  {Dorneich}}]{schmid2002finite}%
  \BibitemOpen
  \bibfield  {author} {\bibinfo {author} {\bibfnamefont {G.}~\bibnamefont
  {Schmid}}, \bibinfo {author} {\bibfnamefont {S.}~\bibnamefont {Todo}},
  \bibinfo {author} {\bibfnamefont {M.}~\bibnamefont {Troyer}},\ and\ \bibinfo
  {author} {\bibfnamefont {A.}~\bibnamefont {Dorneich}},\ }\bibfield  {title}
  {\bibinfo {title} {Finite-temperature phase diagram of hard-core bosons in
  two dimensions},\ }\href {https://doi.org/10.1103/PhysRevLett.88.167208}
  {\bibfield  {journal} {\bibinfo  {journal} {Phys. Rev. Lett.}\ }\textbf
  {\bibinfo {volume} {88}},\ \bibinfo {pages} {167208} (\bibinfo {year}
  {2002})}\BibitemShut {NoStop}%
\bibitem [{\citenamefont {Jordan}\ \emph {et~al.}(2009)\citenamefont {Jordan},
  \citenamefont {Or\'us},\ and\ \citenamefont {Vidal}}]{jordan2009numerical}%
  \BibitemOpen
  \bibfield  {author} {\bibinfo {author} {\bibfnamefont {J.}~\bibnamefont
  {Jordan}}, \bibinfo {author} {\bibfnamefont {R.}~\bibnamefont {Or\'us}},\
  and\ \bibinfo {author} {\bibfnamefont {G.}~\bibnamefont {Vidal}},\ }\bibfield
   {title} {\bibinfo {title} {Numerical study of the hard-core {Bose-Hubbard}
  model on an infinite square lattice},\ }\href
  {https://doi.org/10.1103/PhysRevB.79.174515} {\bibfield  {journal} {\bibinfo
  {journal} {Phys. Rev. B}\ }\textbf {\bibinfo {volume} {79}},\ \bibinfo
  {pages} {174515} (\bibinfo {year} {2009})}\BibitemShut {NoStop}%
\bibitem [{\citenamefont {Hen}\ \emph {et~al.}(2010)\citenamefont {Hen},
  \citenamefont {Iskin},\ and\ \citenamefont {Rigol}}]{hen2010phase}%
  \BibitemOpen
  \bibfield  {author} {\bibinfo {author} {\bibfnamefont {I.}~\bibnamefont
  {Hen}}, \bibinfo {author} {\bibfnamefont {M.}~\bibnamefont {Iskin}},\ and\
  \bibinfo {author} {\bibfnamefont {M.}~\bibnamefont {Rigol}},\ }\bibfield
  {title} {\bibinfo {title} {Phase diagram of the hard-core bose-hubbard model
  on a checkerboard superlattice},\ }\href
  {https://doi.org/10.1103/PhysRevB.81.064503} {\bibfield  {journal} {\bibinfo
  {journal} {Phys. Rev. B}\ }\textbf {\bibinfo {volume} {81}},\ \bibinfo
  {pages} {064503} (\bibinfo {year} {2010})}\BibitemShut {NoStop}%
\bibitem [{Note5()}]{Note5}%
  \BibitemOpen
  \bibinfo {note} {Most physical boson systems have negative hopping
  parameters, corresponding to ferromagnetic in-plane coupling $J^{\perp }<0$
  in the spin model. The antiferromagnetic pseudospin coupling in our system
  leads to frustration and richer phase diagrams that do not exist in most
  boson systems.}\BibitemShut {Stop}%
\bibitem [{\citenamefont {Matsubara}\ and\ \citenamefont
  {Matsuda}(1956)}]{matsubara1956lattice}%
  \BibitemOpen
  \bibfield  {author} {\bibinfo {author} {\bibfnamefont {T.}~\bibnamefont
  {Matsubara}}\ and\ \bibinfo {author} {\bibfnamefont {H.}~\bibnamefont
  {Matsuda}},\ }\bibfield  {title} {\bibinfo {title} {A lattice model of liquid
  helium, {I}},\ }\href@noop {} {\bibfield  {journal} {\bibinfo  {journal}
  {Progress of Theoretical Physics}\ }\textbf {\bibinfo {volume} {16}},\
  \bibinfo {pages} {569} (\bibinfo {year} {1956})}\BibitemShut {NoStop}%
\bibitem [{\citenamefont {Gu}\ \emph {et~al.}(2021)\citenamefont {Gu},
  \citenamefont {Ma}, \citenamefont {Liu}, \citenamefont {Watanabe},
  \citenamefont {Taniguchi}, \citenamefont {Hone}, \citenamefont {Shan},\ and\
  \citenamefont {Mak}}]{gu2021dipolar}%
  \BibitemOpen
  \bibfield  {author} {\bibinfo {author} {\bibfnamefont {J.}~\bibnamefont
  {Gu}}, \bibinfo {author} {\bibfnamefont {L.}~\bibnamefont {Ma}}, \bibinfo
  {author} {\bibfnamefont {S.}~\bibnamefont {Liu}}, \bibinfo {author}
  {\bibfnamefont {K.}~\bibnamefont {Watanabe}}, \bibinfo {author}
  {\bibfnamefont {T.}~\bibnamefont {Taniguchi}}, \bibinfo {author}
  {\bibfnamefont {J.~C.}\ \bibnamefont {Hone}}, \bibinfo {author}
  {\bibfnamefont {J.}~\bibnamefont {Shan}},\ and\ \bibinfo {author}
  {\bibfnamefont {K.~F.}\ \bibnamefont {Mak}},\ }\bibfield  {title} {\bibinfo
  {title} {Dipolar excitonic insulator in a moir\'e lattice},\ }\href@noop {}
  {\bibfield  {journal} {\bibinfo  {journal} {arXiv preprint arXiv:2108.06588}\
  } (\bibinfo {year} {2021})}\BibitemShut {NoStop}%
\bibitem [{\citenamefont {Zhang}\ \emph
  {et~al.}(2021{\natexlab{b}})\citenamefont {Zhang}, \citenamefont {Regan},
  \citenamefont {Wang}, \citenamefont {Zhao}, \citenamefont {Wang},
  \citenamefont {Sayyad}, \citenamefont {Yumigeta}, \citenamefont {Watanabe},
  \citenamefont {Taniguchi}, \citenamefont {Tongay} \emph
  {et~al.}}]{zhang2021correlated}%
  \BibitemOpen
  \bibfield  {author} {\bibinfo {author} {\bibfnamefont {Z.}~\bibnamefont
  {Zhang}}, \bibinfo {author} {\bibfnamefont {E.~C.}\ \bibnamefont {Regan}},
  \bibinfo {author} {\bibfnamefont {D.}~\bibnamefont {Wang}}, \bibinfo {author}
  {\bibfnamefont {W.}~\bibnamefont {Zhao}}, \bibinfo {author} {\bibfnamefont
  {S.}~\bibnamefont {Wang}}, \bibinfo {author} {\bibfnamefont {M.}~\bibnamefont
  {Sayyad}}, \bibinfo {author} {\bibfnamefont {K.}~\bibnamefont {Yumigeta}},
  \bibinfo {author} {\bibfnamefont {K.}~\bibnamefont {Watanabe}}, \bibinfo
  {author} {\bibfnamefont {T.}~\bibnamefont {Taniguchi}}, \bibinfo {author}
  {\bibfnamefont {S.}~\bibnamefont {Tongay}}, \emph {et~al.},\ }\bibfield
  {title} {\bibinfo {title} {{Correlated interlayer exciton insulator in double
  layers of monolayer WSe$_2$ and moir\'e WS$_2$/WSe$_2$}},\ }\href@noop {}
  {\bibfield  {journal} {\bibinfo  {journal} {arXiv preprint arXiv:2108.07131}\
  } (\bibinfo {year} {2021}{\natexlab{b}})}\BibitemShut {NoStop}%
\bibitem [{\citenamefont {Zeng}\ \emph {et~al.}(2022)\citenamefont {Zeng},
  \citenamefont {Xia}, \citenamefont {Dery}, \citenamefont {Watanabe},
  \citenamefont {Taniguchi}, \citenamefont {Shan},\ and\ \citenamefont
  {Mak}}]{zeng2022doublemoire}%
  \BibitemOpen
  \bibfield  {author} {\bibinfo {author} {\bibfnamefont {Y.}~\bibnamefont
  {Zeng}}, \bibinfo {author} {\bibfnamefont {Z.}~\bibnamefont {Xia}}, \bibinfo
  {author} {\bibfnamefont {R.}~\bibnamefont {Dery}}, \bibinfo {author}
  {\bibfnamefont {K.}~\bibnamefont {Watanabe}}, \bibinfo {author}
  {\bibfnamefont {T.}~\bibnamefont {Taniguchi}}, \bibinfo {author}
  {\bibfnamefont {J.}~\bibnamefont {Shan}},\ and\ \bibinfo {author}
  {\bibfnamefont {K.~F.}\ \bibnamefont {Mak}},\ }\href
  {https://doi.org/10.48550/ARXIV.2205.07354} {\bibinfo {title} {Exciton
  density waves in coulomb-coupled dual moiré lattices}} (\bibinfo {year}
  {2022})\BibitemShut {NoStop}%
\bibitem [{\citenamefont {Kang}\ and\ \citenamefont
  {Vafek}(2020)}]{kang2020non}%
  \BibitemOpen
  \bibfield  {author} {\bibinfo {author} {\bibfnamefont {J.}~\bibnamefont
  {Kang}}\ and\ \bibinfo {author} {\bibfnamefont {O.}~\bibnamefont {Vafek}},\
  }\bibfield  {title} {\bibinfo {title} {Non-abelian dirac node braiding and
  near-degeneracy of correlated phases at odd integer filling in magic-angle
  twisted bilayer graphene},\ }\href
  {https://doi.org/10.1103/PhysRevB.102.035161} {\bibfield  {journal} {\bibinfo
   {journal} {Phys. Rev. B}\ }\textbf {\bibinfo {volume} {102}},\ \bibinfo
  {pages} {035161} (\bibinfo {year} {2020})}\BibitemShut {NoStop}%
\bibitem [{\citenamefont {Xu}\ \emph {et~al.}(2022)\citenamefont {Xu},
  \citenamefont {Kang}, \citenamefont {Watanabe}, \citenamefont {Taniguchi},
  \citenamefont {Mak},\ and\ \citenamefont {Shan}}]{xu2022tunable}%
  \BibitemOpen
  \bibfield  {author} {\bibinfo {author} {\bibfnamefont {Y.}~\bibnamefont
  {Xu}}, \bibinfo {author} {\bibfnamefont {K.}~\bibnamefont {Kang}}, \bibinfo
  {author} {\bibfnamefont {K.}~\bibnamefont {Watanabe}}, \bibinfo {author}
  {\bibfnamefont {T.}~\bibnamefont {Taniguchi}}, \bibinfo {author}
  {\bibfnamefont {K.~F.}\ \bibnamefont {Mak}},\ and\ \bibinfo {author}
  {\bibfnamefont {J.}~\bibnamefont {Shan}},\ }\bibfield  {title} {\bibinfo
  {title} {{Tunable bilayer Hubbard model physics in twisted WSe$_2$}},\
  }\href@noop {} {\bibfield  {journal} {\bibinfo  {journal} {arXiv preprint
  arXiv:2202.02055}\ } (\bibinfo {year} {2022})}\BibitemShut {NoStop}%
\end{thebibliography}%

\end{document}